\documentstyle[12pt]{article}
\newlength{\extraspace}
\newlength{\extraspaces}
\newcommand{\ba}{\begin{eqnarray}
\addtolength{\abovedisplayskip}{\extraspaces}
\addtolength{\belowdisplayskip}{\extraspaces}
\addtolength{\abovedisplayshortskip}{\extraspace}
\addtolength{\belowdisplayshortskip}{\extraspace}}
\newcommand{\ea}{\end{eqnarray}}

\newcommand{\nonu}{\nonumber \\[.5mm]}
\newcommand{\A}{&\!\!\!}

\begin{document}

\thispagestyle{empty}

\hfill \parbox{3.5cm}{SIT-LP-01/04 \\ hep-th/0109042}
\vspace*{1cm}
\vspace*{1cm}
\begin{center}
{\bf ON STRUCTURE OF SUPERON-GRAVITON MODEL(SGM)} \\[20mm]
{Kazunari SHIMA and Motomu TSUDA} \\[2mm]
{\em Laboratory of Physics,  Saitama Institute of Technology}\footnote{e-mail: shima@sit.ac.jp, tsuda@sit.ac.jp}\\
{\em Okabe-machi, Saitama 369-0293, Japan}\\[2mm]
{August  2001}\\[15mm]


\begin{abstract}
The fundamental action of superon-graviton model(SGM) for space-time and matter 
is written down explicitly in terms of the fields of the graviton and superons 
by using the affine and the spin connection formalisms, alternatively.  
Some  characteristic structures including some hidden symmetries of the gravitational 
coupling of superons are manifested (in two dimensional space-time) with some details of 
the calculations. 
SGM cosmology is discussed  briefly.

PACS:04.50.+h, 12.60.Jv, 12.60.Rc, 12.10.-g /Keywords: supersymmetry, graviton, 
Nambu-Goldstone fermion, unified theory
\end{abstract}
\end{center}

\newpage
{\bf 1. Introduction}             \\
It seems inevitable to introduce  new particles yet to be observed and  new (gauge) symmetries 
for  exploring the new physics and the new framework for the unification of space-time and matter 
beyond the standard model(SM). 
Supersymmetry(SUSY)\cite{wzgl} \cite{va} may be the most promissing gauge symmetry  beyond SM, 
especially for the unification of space-time and matter.      
Nambu-Goldstone(N-G) fermion\cite{va}\cite{o} would appear in the spontaneous SUSY breaking and 
plays essentially important roles in the unified model building.                                   \\
Here it is useful to distinguish the qualitative differences of the origins of N-G fermion.
In O'Raifeartaigh model N-G fermion stems from the symmetry of the dynamics(interaction) of the linear 
representation (multiplet) of SUSY, 
while in Volkov-Akulov model N-G fermion stems from the symmetry of space-time and gives the nonlinear(NL) 
representation of SUSY\cite{ks0}. 
As demonstrated in supergravity(SUGRA) coupled with Volkov-Akulov model it is rather well understood 
in the linear realization of  SUSY(L SUSY) that N-G fermion is 
converted to the longitudinal component of spin 3/2 field by the super-Higgs mechanism 
and  breaks local linear SUSY spontaneously by giving mass to gravitino\cite{dz}. 
N-G fermion degrees of freedom become unphysical in the low energy, so far.                          \\
While, speculating that the chirality ( ,i.e. the massless fermionic state ) of SM may indicate 
the N-G fermion nature of matter behind SM, it may be interesting to servey other possibilities 
concerning how SUSY is realized and where N-G fermion has gone (in the low energy).                  \\
In the previous paper \cite{ks1} we have introduced a new fundamental constituent with spin 1/2 
${\em superon}$ and  proposed ${\em superon-graviton\ model}$(SGM) equippted with NL SUSY
as a model for unity of space-time and matter. 
In SGM, the fundamental entities of nature are 
the graviton with spin-2 and a  quintet of superons  with spin-1/2.  
They are the elementary gauge fields corresponding to the ordinary local GL(4,R) and 
the global nonlinear supersymmetry(NL SUSY) with a global SO(10), i.e. N=10 V-A model, respectively. 
Interestingly, the quantum numbers of the superon-quintet are the same as those of the fundamental 
representation ${\underline 5}$ of the matter multiplet of SU(5) GUT\cite{gg}.
All observed elementary particles including gravity are assigned to a single  irreducible 
massless representation of SO(10) super-Poincar\'e(SP) symmetry and reveals a remarkable potential  
for the phenomenology, e.g. they may explain naturally the three-generations structure of quarks and leptons, 
the stability of proton, various mixings, ..etc\cite{ks1}. 
And in SGM except graviton they are supposed to be the (massless) eigenstates of superons of 
SO(10) SP symmetry \cite{ks2} of space-time and matter. The uniqueness of N=10 among all SO(N) SP is pointed out. 
The arguments are group theoretical so far.  \\
In order to obtain the fundamental action 
of SGM which is invariant at least under local GL(4,R), local Lorentz, global NL SUSY transformations 
and global SO(10), 
we have performed the similar geometrical arguments to Einstein genaral relativity theory(EGRT) 
in the SGM space-time, 
where the tangent (Riemann-flat) Minkowski space-time is specified by the coset space SL(2,C) coordinates 
(corresponding to Nambu-Goldstone(N-G) fermion) of NL SUSY of Volkov-Akulov(V-A)\cite{va} 
in addition to the ordinary Lorentz SO(3,1) coordinates\cite{ks1}, which are locally homomorphic groups\cite{ks3}. 
As shown in Ref.\cite{ks3} the  SGM  action for the unified  SGM space-time 
is naturally the analogue of Einstein-Hilbert(E-H) action of GR 
and has the similar concise expression.  And interestingly it may be regarded as a kind of a generalization 
of Born-Infeld action\cite{bi}. (The similar systematic arguments are applicable to 
spin 3/2 N-G case.\cite{st1})                           \\
In this article, after a  brief review of SGM for the self contained arguments 
we expand SGM  action in terms of the fields of graviton and superons in order to 
see some characteristic structures of our model and also show some details of the calculations.     \\
For the sake of the comparison the expansion is performed by the affine connection 
formalism and by the spin connection formalism.           \\
Finally some hidden symmetries and a potential cosmology, especially the birth of the universe 
are mentioned briefly. 

{\bf 2. Fundamental action of superon-graviton model(SGM)}                  \\
In Ref.\cite{ks3}, SGM space-time is defined as the space-time whose tangent(flat) space-time is specified by 
SO(1,3) Lorentz coordinates ${x^{a}}$  and  the coset space SL(2,C) coordinates ${\psi}$ of  NL SUSY of 
Volkov-Akulov(V-A)\cite{va}. 
The  unified  vierbein  ${w_{a}}^{\mu}$ and the unified metric 
$s^{\mu\nu}(x) \equiv w{_a}^{\mu}(x) w^{a \nu}(x)$ of SGM space-time 
are defined by generalizing the NL SUSY invariant differential forms of V-A to the curved space-time\cite{ks3}. 
SGM action is given as follows\cite{ks3}
\begin{equation}
L_{SGM}=-{c^{3} \over 16{\pi}G}\vert w \vert(\Omega + \Lambda ),
\label{SGMac}
\end{equation}
\begin{equation}
\vert w \vert=det{w_{a}}^{\mu}=det({e_{a}}^{\mu}+ {t_{a}}^{\mu}),  \quad
{t_{a}}^{\mu}={\kappa  \over 2i}\sum_{j=1}^{10}(\bar{\psi}^{j}\gamma_{a}
\partial^{\mu}{\psi}^{j}
- \partial^{\mu}{\bar{\psi}^{j}}\gamma_{a}{\psi}^{j}), 
\label{SGMtw}
\end{equation}
where $\kappa(=\kappa_{V-A})$ is an arbitrary constant up now with the dimension of the fourth power of length, 
${e_{a}}^{\mu}$ and ${\psi}^{j}(j=1,2,..10)$ are the fundamental elememtary fields of SGM, i.e. 
the vierbein of Einstein general relativity theory(EGRT) 
and the superons of N-G fermion of NL SUSY of Volkov-Akulov\cite{va}, respectively. 
$\Lambda$ is a cosmological constant which is necessary for SGM action to reduce to 
V-A model with the first order derivative terms  of the superon in the Riemann-flat space-time. 
$\Omega$ is a unified   scalar curvature of SGM space-time
analogous to the Ricci scalar curvature $R$ of EGRT. 
SGM action (\ref{SGMac}) is invariant under the following new SUSY transformations
\begin{equation}
\delta \psi^{i}(x) = \zeta^{i} + i \kappa (\bar{\zeta}^{j}{\gamma}^{\rho}\psi^{j}(x)) \partial_{\rho}\psi^{i}(x),
\label{newSUSY1/2}
\end{equation} 
\begin{equation}
\delta {e^{a}}_{\mu}(x) = i \kappa (\bar{\zeta}^{j}{\gamma}^{\rho}\psi^{j}(x))D_{[\rho} {e^{a}}_{\mu]}(x),
\label{newSUSY2}
\end{equation} 
where $\zeta^{i}, (i=1,..10)$ is a constant spinor parameter, 
$D_{[\rho} {e^{a}}_{\mu]}(x) = D_{\rho}{e^{a}}_{\mu}-D_{\mu}{e^{a}}_{\rho}$ and $D_{\mu}$ is a covariant derivative 
containing a symmetric affine connection. 
The explicit expression of $\Omega$ is obtained 
by just replacing $e{_a}^{\mu}(x)$ in Ricci scalar $R$ 
of EGRT by the unified vierbein $w{_a}^{\mu}(x) = e{_a}^{\mu} + t{_a}^{\mu}$ 
of the SGM curved space-time, which gives the gravitational interaction of 
$\psi(x)$ invariant under (\ref{newSUSY1/2}) and  (\ref{newSUSY2}). 
The invariance can be easily undestood by observing that under (\ref{newSUSY1/2}) and  (\ref{newSUSY2}) 
the new vierbein $w{_a}^{\mu}(x)$ and the new metric $s_{\mu\nu}(x)$ have general coordinate transformations
\cite{ks3}. 
\ba
\A \A
\delta_{\zeta} {w^{a}}_{\mu} = \xi^{\nu} \partial_{\nu}{w^{a}}_{\mu} + \partial_{\mu} \xi^{\nu} {w^{a}}_{\nu}, 
\label{gl4r-w}
\ea
\ba
\A \A \delta_{\zeta} s_{\mu\nu} = \xi^{\kappa} \partial_{\kappa}s_{\mu\nu} +  
\partial_{\mu} \xi^{\kappa} s_{\kappa\nu} 
+ \partial_{\nu} \xi^{\kappa} s_{\mu\kappa}, 
\label{gl4r-s}
\ea
where  $\xi^{\rho}=i \kappa (\bar{\zeta}^{j}{\gamma}^{\rho}\psi^{j}(x))$.
The overall factor of SGM action  is fixed  to ${-c^3 \over 16{\pi}G}$, 
which reproduces E-H action of GR in the absence of superons(matter). 
Also in the Riemann-flat space-time, i.e.  $e{_a}^{\mu}(x) \rightarrow \delta{_a}^{\mu}$,  
it reproduce V-A action of NL SUSY\cite{va} with ${{\kappa}^{-1}}_{V-A} = 
{c^3 \over 16{\pi}G}{\Lambda}$ in the first order derivative terms  of the superon. 
Therefore our model(SGM) predicts a (small) non-zero cosmological constant, provided 
${\kappa}_{V-A} \sim O(1)$,  and posesses two mass scales. 
Furthermore it fixes the coupling constant of superon(N-G fermion) with the vacuum to 
$({c^3 \over 16{\pi}G}{\Lambda})^{1 \over 2}$ (from the low energy theorem viewpoint), 
which may be relevant to the birth (of the matter and Riemann space-time) of the universe.    \\
It is interesting that our action is the vacuum (matter free) action in  SGM space-time 
as read off from (\ref{SGMac}) but gives in ordinary Riemann space-time the E-H action with matter(superons) 
accompanying the spontaneous supersymmetry breaking.   \\
The commutators of new SUSY transformations induces the generalized general coordinate transformations 
\ba
\A \A [\delta_{\zeta_1}, \delta_{\zeta_2}] \psi
      = \Xi^{\mu} \partial_{\mu} \psi,
\label{com1/2-p} \\
\A \A [\delta_{\zeta_1}, \delta_{\zeta_2}] e{^a}_{\mu}
      = \Xi^{\rho} \partial_{\rho} e{^a}_{\mu}
      + e{^a}_{\rho} \partial_{\mu} \Xi^{\rho},
\label{com1/2-e}
\ea
where $\Xi^{\mu}$ is defined by
\begin{equation}
\Xi^{\mu} = 2ia (\bar{\zeta}_2 \gamma^{\mu} \zeta_1)
      - \xi_1^{\rho} \xi_2^{\sigma} e{_a}^{\mu}
      (D_{[\rho} e{^a}_{\sigma]}).
\end{equation}
We have shown that our action is invariant at least under\cite{st2} 
\begin{equation}
[{\rm global\ NL\ SUSY}] \otimes [{\rm local\ GL(4,R)}]
\otimes [{\rm local\ Lorentz}] \otimes [{\rm global\ SO(N)}], 
\label{SGMgr}
\end{equation}
which is isomorphic to N=10 extended (global SO(10)) SP symmetry through which SGM 
reveals the spectrum of all observed particles in the low energy\cite{ks2}. 
In contrast with the ordinary SP SUSY, new SUSY may be regarded as a square root of 
a generalized  GL(4,R). The usual local GL(4,R) invariance is obvious by the construction.  \\
The simple expression (\ref{SGMac}) invariant under the above symmetry   may be universal for 
the gravitational coupling of Nambu-Goldstone(N-G) fermion, for by performing the parallel arguments 
we obtain the same expression for the gravitational interaction of the spin-3/2 N-G fermion\cite{st1}.   \\
Now  to clarify the characteristic features of SGM 
we focus on  N=1 SGM  for simplicity  without loss of generality and 
write down the action explicitly in terms of ${t^{a}}_{\mu}$(or $\psi$) 
and $g^{\mu\nu}$(or ${e^{a}}_{\mu}$). We will see that the graviton and superons(matter) are complementary 
in SGM and contribute equally to the curvature of SGM space-time. 
Contrary to its  simple expression (\ref{SGMac}), it has rather complicated and rich structures.   \\
We use the  Minkowski tangent space metric 
${1 \over 2}\{ \gamma^a, \gamma^b \} = \eta^{ab}= (+, -, -, -)$
and $\sigma^{ab} = {i \over 4}[\gamma^a, \gamma^b]$. 
(Latin $(a,b,..)$ and Greek $(\mu,\nu,..)$ are the indices
for local Lorentz and general coordinates, respectively.)
By requiring that the unified action of SGM space-time should reduce to V-A in the  flat space-time 
which is specified by $x^{a}$ and $\psi(x)$ and that the graviton and superons contribute equally to the 
unified curvature of SGM space-time,  it is natural to consider that 
the unified vierbein  $w{^a}_{\mu}(x)$ and the unified metric $s^{\mu\nu}(x)$ of unified SGM space-time 
are defined  through the NL SUSY invariant differential
forms $\omega^a$ of V-A\cite{va} as follows: 

\ba
\A \A \omega^a = w{^a}_{\mu} dx^{\mu},
\label{om} \\
\A \A w{^a}_{\mu}(x) = e{^a}_{\mu}(x) + t{^a}_{\mu}(x),
\label{new-w}
\ea
where $e{^a}_{\mu}(x)$ is the vierbein of EGRT and $t{^a}_{\mu}(x)$
is defined by
\begin{equation}
t{^a}_{\mu}(x) = i\kappa \bar{\psi}\gamma^{a}
\partial_{\mu}{\psi},
\label{t-1/2}
\end{equation}
where the first and the second indices of ${t^{a}}_{\mu}$ represent those of the $\gamma$ matrices and 
the general covariant derivatives, respectively.
We can easily obtain the inverse $w{_a}^{\mu}$ of the  vierbein
$w{^a}_{\mu}$ in the power series of $t{^a}_{\mu}$ as follows, 
which terminates with $t^4$(for 4 dimensional space-time):
\begin{equation}
w{_a}^{\mu} = e{_a}^{\mu}
- t{^{\mu}}_a + t{^{\rho}}_a t{^{\mu}}_{\rho} - t{^{\rho}}_a t{^{\sigma}}_{\rho} t{^{\mu}}_{\sigma} 
+ t{^{\rho}}_a t{^{\sigma}}_{\rho} t{^{\kappa}}_{\sigma}t{^{\mu}}_{\kappa}.
\label{new-wi}
\end{equation}
Similarly a new metric tensor $s_{\mu\nu}(x)$ and its inverse $s^{\mu\nu}(x)$ 
are  introduced in SGM curved space-time as follows:
\ba
\A \A s_{\mu\nu}(x) \equiv w{^a}_{\mu}(x) w_{a \nu}(x) =w{^a}_{\mu}(x) \eta_{ab} w{^b}_{ \nu}(x) \nonu
\A \A \hspace{1.5cm}
= g_{\mu\nu} + t_{\mu\nu} + t_{\nu\mu} + {t^{\rho}}_{\mu} {t_{\rho\nu}}.
\label{new-s}
\ea
and 
\ba
\A \A s^{\mu\nu}(x) \equiv w{_a}^{\mu}(x) w^{a \nu}(x) \nonu
\A \A \hspace{1.0cm}     
      =  g^{\mu\nu}  \nonu
\A \A \hspace{1.0cm}      
      - t^{\mu\nu} - t^{\nu\mu} \nonu
\A \A \hspace{1.0cm}     
      + t^{\rho\mu}{t^{\nu}}_{\rho} + t^{\rho\nu}{t^{\mu}}_{\rho} +  t^{\mu\rho}{t^{\nu}}_{\rho}  \nonu
\A \A \hspace{1.0cm}     
      - t^{\rho\mu}{t^{\sigma}}_{\rho} {t^{\nu}}_{\sigma} - t^{\rho\nu}{t^{\sigma}}_{\rho}{t^{\mu}}_{\rho}  
      - t^{\mu\sigma}{t^{\rho}}_{\sigma}{t^{\nu}}_{\rho} - t^{\nu\rho}{t^{\sigma}}_{\rho}{t^{\mu}}_{\sigma} \nonu
\A \A \hspace{1.0cm}      
      + t^{\rho\mu}{t^{\sigma}}_{\rho}{t^{\kappa}}_{\sigma}{t^{\nu}}_{\kappa} 
      + t^{\rho\nu}{t^{\sigma}}_{\rho}{t^{\kappa}}_{\sigma}{t^{\mu}}_{\kappa}         \nonu
\A \A \hspace{1.0cm}      
      + t^{\mu\sigma}{t^{\rho}}_{\sigma}{t^{\sigma}}_{\rho}{t^{\nu}}_{\sigma} 
      + t^{\nu\sigma}{t^{\rho}}_{\sigma}{t^{\sigma}}_{\rho}{t^{\mu}}_{\sigma} 
      + t^{\rho\kappa}{t^{\sigma}}_{\kappa} {t^{\mu}}_{\rho}{t^{\nu}}_{\sigma}. 
\label{new-si}
\ea
We can easily show 
\begin{equation}
{w_a}^{\mu} w_{b \mu} = \eta_{ab}, s_{\mu \nu}{w_a}^{\mu} {w_b}^{\nu} = \eta_{ab}.
\label{new-metric}
\end{equation} 
It is obvious from the above general covariant arguments that (\ref{SGMac}) is invariant under the ordinaly GL(4,R) 
and under (\ref{newSUSY1/2}) and (\ref{newSUSY2}).     \\
By using (\ref{new-w}), (\ref{new-wi}), (\ref{new-s}) and (\ref{new-si}) we can express  
SGM action  (\ref{SGMac}) in terms of  $e{^a}_{\mu}(x)$ and  $\psi^{j}(x)$, which describes explicitly 
the fundamental interaction of graviton with  superons. 
The expansion  of the action in terms of the power series of $\kappa$ (or ${t^{a}}_{\mu}$) 
can be carried out straightforwardly.  
After  the lengthy  calculations concerning the complicated  structures of the indices 
we obtain       \\   
\ba
\A \A L_{SGM} = - {c^3\Lambda \over 16{\pi}G} e \vert w_{V-A} \vert - {c^3 \over 16{\pi}G} e R  \nonu
\A \A \hspace{1.5cm}
+ {c^3 \over 16{\pi}G} e \big[ \ 2 t^{(\mu\nu)} R_{\mu\nu}  \nonu
\A \A \hspace{1.5cm}
+ {1 \over 2} \{ g^{\mu\nu}\partial^{\rho}\partial_{\rho}t_{(\mu\nu)}
- t_{(\mu\nu)}\partial^{\rho}\partial_{\rho}g^{\mu\nu}       \nonu
\A \A \hspace{1.5cm}
+ g^{\mu\nu}\partial^{\rho}t_{(\mu\sigma)}\partial^{\sigma}g_{\rho\nu}
- 2g^{\mu\nu}\partial^{\rho}t_{(\mu\nu)}\partial^{\sigma}g_{\rho\sigma}
- g^{\mu\nu}g^{\rho\sigma}\partial^{\kappa}t_{(\rho\sigma)}\partial^{\kappa}g_{\mu\nu} \}     \nonu
\A \A \hspace{1.5cm}
+ ({t^{\mu}}_{\rho}t^{\rho\nu}+{t^{\nu}}_{\rho}t^{\rho\mu}+t^{\mu\rho}{t^{\nu}}_{\rho})R_{\beta\mu}   \nonu
\A \A \hspace{1.5cm}
- \{ 2 t^{(\mu\rho)}{t^{(\nu}}_{\rho)}R_{\mu\nu} + t^{(\mu \rho)} t^{(\nu \sigma)} R_{\mu \nu \rho \sigma} \nonu
\A \A \hspace{1.5cm}
+ {1 \over 2}t^{(\mu\nu)}( g^{\rho\sigma}\partial^{\mu}\partial_{\nu}t_{(\rho\sigma)} 
- g^{\rho\sigma}\partial^{\rho}\partial_{\mu}t_{(\sigma\nu)} + \dots )  \}  \nonu
\A \A \hspace{1.5cm}
+\{ O(t^{3})\} +\{ O(t^{4})\} + \dots + \{ O(t^{10})\}) \big],
\label{L-exp}
\ea
where $e=det{e^{a}}_{\mu}$, $t^{(\mu\nu)}=t^{\mu\nu}+t^{\nu\mu}$, $t_{(\mu\nu)}=t_{\mu\nu}+t_{\nu\mu}$, and 
$ \vert w_{V-A} \vert = det{w^{a}}_{b} $ is the flat space V-A action\cite{va} 
containing up to $O(t^{4})$ and $R$ and $R_{\mu\nu}$ are the Ricci curvature tensors  of GR.    \\
Remarkably the first term can be regarded as a space-time dependent cosmological term
and  reduces to V-A action \cite{va} with ${\kappa_{V-A}}^{-1} = {c^3 \over 16{\pi}G}{\Lambda}$ 
in the Riemann-flat $e{_a}^{\mu}(x) \rightarrow \delta{_a}^{\mu}$ space-time. 
The second term is the familiar E-H action of GR. 
These expansions show the complementary relation of graviton and (the stress-energy tensor of) superons.  
The existence of (in the Riemann-flat space-time) NL SUSY invariant terms  
with the (second order) derivatives of the superons beyond V-A model are manifested. 
For example, the lowest order of such terms appear in $O(t^{2})$ and have the following expressions 
(up to the total derivative terms)                                           \\ 
\begin{equation}
+\epsilon^{abcd}{\epsilon_{a}}^{efg}\partial_{c}t_{(be)}\partial_{f}t_{(dg)}.
\label{b-va}
\end{equation}
The existence of such  derivative  terms 
in addition to the original V-A model are already pointed out and exemplified  in part in \cite{sw}.  
Note that (\ref{b-va}) vanishes in 2 dimensional space-time.                                  \\
Here we just mention that we can consider two  types of the flat space in SGM, which are not equivalent. 
One is SGM-flat, i.e. ${w_{a}}^{\mu}(x) \rightarrow {\delta_{a}}^{\mu}$,  space-time and 
the other is Riemann-flat, i.e.  $e{_a}^{\mu}(x) \rightarrow \delta{_a}^{\mu}$, space-time, 
where SGM action reduces to  ${-{c^3\Lambda \over 16{\pi}G}}$ and                                
${-{c^3\Lambda \over 16{\pi}G} \vert w_{V-A} \vert - 
{c^3 \over 16{\pi}G}( derivative \  terms) }$, respectively. 
Note that SGM-flat space-time may allow Riemann space-time, 
e.g. $t{_a}^{\mu}(x) \rightarrow  -e{_a}^{\mu} + \delta{_a}^{\mu}$ realizes 
Riemann  space-time $and$ SGM-flat space-time. The cosmological implications are mentioned  in the 
discussions.                \\  

{\bf 3. SGM in two  dimensional space-time}    \\ 
It is well known that two dimensional GR has no physical degrees of freedom ( due to the local GL(2,R)). 
SGM in SGM space-time is also the case.
However the the arguments with the general covariance shed light on the  characteristic off-shell gauge structures 
of the theory in any space-time dimensions. 
Especialy for SGM, it is also useful for linearlizing the theory  to see explicitly the superon-graviton 
coupling in (two dimensional) Riemann space-time.  The expansion  holds  up to $O(t^{2})$ in four dimensional 
spacetime as well.  \\
{\bf 3.1 SGM in affine connection formalism}     \\
Now we go to two dimensional SGM space-time to simplify the arguments without loss of generality and 
demonstrate some details of the computations. 
We adopt firstly  the affine connection formalism. 
The knowledge of the complete  structure of SGM action including the surface terms 
is  useful to linearlize SGM into the equivalent linear theory and to find the symmetry breaking of the model.   \\
Following EGRT the scalar curvature tensor $\Omega$ of SGM space-time is given as follows 
\ba
\Omega = \A \A s^{\beta\mu}\Omega{^{\alpha}}_{\beta\mu \alpha}  \nonu
\A \A =s^{\beta\mu} [\{ \partial_{\mu}{\Gamma^{\lambda}}_{\beta\alpha} 
       +{\Gamma^{\alpha}}_{\lambda\mu}{\Gamma^{\lambda}}_{\beta\alpha} \} 
       -\{ \ lower \ indices  (\mu \leftrightarrow \alpha)  \}],      
\label{Omega}
\ea
where the Christoffel symbol of the second kind of SGM space-time is 
\ba
\A \A {\Gamma^{\alpha}}_{\beta\mu}
={1 \over 2}s^{\alpha\rho}\{ \partial_{\beta}s_{\rho\mu}+\partial_{\mu}s_{\beta\rho}-\partial_{\rho}s_{\mu\beta}\}. 
\label{affine}
\ea
The straightforwad expression of SGM action (\ref{SGMac}) in two dimensional space-time, (which is $3^{6}$ times 
more complicated than the two dimensional GR), is given as follows     \\
\ba
\A \A L_{2dSGM} = - {c^3 \over 16{\pi}G} e \{ 1+{t^{a}}_{a}    
+{1 \over 2}({t^{a}}_{a}{t^{b}}_{b}-{t^{a}}_{b}{t^{b}}_{a}) \}  
(g^{\beta\mu}- {\tilde t}^{(\beta\mu)} + {\tilde t}^{2(\beta\mu)})    \nonu
\A \A \hspace{1.5cm}
\times [ \{ {1 \over 2}\partial_{\mu}(g^{\alpha\sigma}- \tilde t^{(\alpha\sigma)} + {\tilde t}^{2(\alpha\sigma)}) 
\partial_{\dot \beta}(g_{\dot \sigma \dot \alpha} +{{\b t}}_{(\dot\sigma \dot\alpha)}+
{{ {\b t}}^{2}}_{(\dot\sigma \dot\alpha)})    \nonu
\A \A \hspace{1.5cm}
+ {1 \over 2}(g^{\alpha\sigma}- \tilde t^{(\alpha\sigma)} + {\tilde t}^{2(\alpha\sigma)})
\partial_{\mu}\partial_{\dot \beta}
(g_{\dot \sigma \dot \alpha} + {\b t}_{(\dot\sigma \dot\alpha)}+{ {\b t}^{2}}_{(\dot\sigma \dot\alpha)} \}    \nonu
\A \A \hspace{1.5cm}
-\{ lower \ indices  (\mu \leftrightarrow \alpha) \}        \nonu
\A \A \hspace{1.5cm}
+\{ {1 \over 4}(g^{\alpha\sigma}- {\tilde t}^{(\alpha\sigma)} + {\tilde t}^{2(\alpha\sigma)})
\partial_{\dot \lambda}(g_{\dot\sigma \dot\mu} + {\b t}_{(\dot\sigma \dot\mu)}+
{{\b t}^{2}}_{(\dot\sigma \dot\mu)})  \nonu
\A \A \hspace{1.5cm} 
(g^{\lambda\rho}- {\tilde t}^{(\lambda\rho)} + {\tilde t}^{2(\lambda\rho)})      
\partial_{\dot \beta}(g_{\dot\rho \dot\alpha} + {\b t}_{(\dot\rho \dot\alpha)}+
{ {\b t}^{2}}_{(\dot\rho \dot\alpha)}) \}    \nonu
\A \A \hspace{1.5cm} 
- \{ lower \ indices  (\mu \leftrightarrow \alpha)  \}]   \nonu
\A \A \hspace{1.5cm}
- {c^3\Lambda \over 16{\pi}G} e \vert w_{V-A} \vert,     \nonu
\A \A \hspace{1.5cm}
\label{2dSGM}
\ea
where we have put 
\ba
\A \A s_{\alpha\beta}=g_{\alpha\beta}+{\b t}_{(\alpha\beta)}+{{\b t}^{2}}_{(\alpha\beta)}, 
s^{\alpha\beta}=g^{\alpha\beta}-{\tilde t}^{(\alpha\beta)}+{\tilde t}^{2(\alpha\beta)},       \nonu
\A \A \hspace{1.5cm}
{\b t}_{(\mu\nu)}=t_{\mu\nu}+t_{\nu\mu}, {{\b t}^{2}}_{(\mu\nu)}={t^{\rho}}_{\mu}t_{\rho\nu},    \nonu
\A \A \hspace{1.5cm} 
{\tilde t}^{(\mu\nu)}=t^{\mu\nu}+t^{\nu\mu},  
{\tilde t}^{2(\mu\nu)}={t^{\mu}}_{\rho}t^{\rho\nu}+{t^{\nu}}_{\rho}t^{\rho\mu}+t^{\mu\rho}{t^{\nu}}_{\rho},    \nonu
\A \A \hspace{1.5cm}   
\label{2dSGM-not1}
\ea
and  the Christoffel symbols of the first kind of SGM space-time contained in (\ref{affine}) are abbreviated as 
\ba
\A \A \partial_{\dot \mu}{ g}_{\dot \sigma \dot \nu} =\partial_{\mu}{g}_{\sigma\nu}
+\partial_{\nu}{g}_{\mu\sigma}-\partial_{\sigma}{ g}_{\nu\mu},                         \nonu
\A \A \hspace{1.5cm}   
\partial_{\dot \mu}{\b t}_{\dot \sigma \dot \nu} =\partial_{\mu}{\b t}_{(\sigma\nu)}
+\partial_{\nu}{\b t}_{(\mu\sigma)}-\partial_{\sigma}{\b t}_{(\nu\mu)},             \nonu
\A \A \hspace{1.5cm}   
\partial_{\dot\mu}{{\b t}^{2}}_{\dot\sigma \dot\nu} =\partial_{\mu}{{\b t}^{2}}_{(\sigma\nu)}+
\partial_{\nu}{{\b t}^{2}}_{(\mu\sigma)}-\partial_{\sigma}{{\b t}^{2}}_{(\nu\mu)}.  \nonu
\A \A \hspace{1.5cm}
\label{2dSGM-not2}
\ea
By  expanding the scalar curvature  $\Omega$ in the power series of $t$ 
which terminates with $t^{4}$, we have the following complete expression of  two dimensional SGM,   \\ 
\ba
\A \A L_{2dSGM} = - {c^3\Lambda \over 16{\pi}G} e \vert w_{V-A} \vert                    \nonu
\A \A \hspace{1.5cm}
- {c^3 \over 16{\pi}G} e \vert w_{V-A} \vert  \big[  R                         \nonu
\A \A \hspace{1.5cm} 
- 2{\tilde t}^{(\mu\nu)} R_{\mu\nu}         \nonu
\A \A \hspace{1.5cm}
+ {1 \over 2} \{ g^{\mu\nu}\partial^{\rho}\partial_{\rho}{\b t}_{(\mu\nu)}
- {\b t}^{(\mu\nu)}\partial^{\rho}\partial_{\rho}g_{\mu\nu}       \nonu
\A \A \hspace{1.5cm}
+ g^{\mu\nu}\partial^{\rho}{\b t}_{(\mu\sigma)}\partial^{\sigma}g_{\rho\nu}
- 2g^{\mu\nu}\partial^{\rho}{\b t}_{(\mu\nu)}\partial^{\sigma}g_{\rho\sigma}
- g^{\mu\nu}g^{\rho\sigma}\partial^{\kappa}{\b t}_{(\rho\sigma)}\partial_{\kappa}g_{\mu\nu} \}  \nonu
\A \A \hspace{1.5cm} 
+ {\tilde t}^{2(\beta\mu)}R_{\beta\mu}         \nonu
\A \A \hspace{1.5cm} 
+ {\tilde t}^{(\beta\mu)}{\tilde t}^{(\alpha\sigma)}R_{\mu\alpha\sigma\beta}        \nonu
\A \A \hspace{1.5cm} 
- {1 \over 2}{\tilde t}^{(\beta\mu)}  \{ g^{\alpha\sigma}\partial_{\mu}\partial_{\beta}{\b t}_{(\alpha\sigma)}  
- \partial^{\sigma}\partial_{\beta}{\b t}_{(\sigma\mu)} 
+ \partial_{\mu}{\tilde t}^{(\alpha\sigma)} \partial_{\beta}g_{\sigma\alpha}
- \partial_{\mu}g^{\alpha\sigma}\partial_{\beta}{\b t}_{(\sigma\alpha)}                            \nonu
\A \A \hspace{1.5cm} 
+ \partial_{\alpha}g^{\alpha\sigma}\partial_{\beta}{\b t}_{(\sigma\mu)} 
- \partial_{\alpha}{\tilde t}^{(\alpha\sigma)} \partial_{\beta}g_{\sigma\mu}                     
+ 2\partial^{\rho}{\b t}_{(\sigma\mu)} \partial^{\sigma}g_{\beta\rho}              \nonu
\A \A \hspace{1.5cm} 
- 2g^{\alpha\sigma}\partial_{\lambda}{\b t}_{(\sigma\mu)} \partial^{\lambda}g_{\alpha\beta}         \nonu
\A \A \hspace{1.5cm} 
+ g^{\alpha\sigma}g^{\lambda\rho} \partial_{\mu}{\b t}_{(\lambda\sigma)} \partial^{\beta}g_{\rho\alpha} 
- 2g^{\alpha\sigma}\partial^{\rho}{\b t}_{(\sigma\alpha)} \partial_{\beta}g_{\rho\mu}         
+ g^{\alpha\sigma}\partial_{\lambda}{\b t}_{(\sigma\alpha)} \partial^{\lambda}g_{\mu\beta}  \}         \nonu
\A \A \hspace{1.5cm} 
-g^{\beta\mu}\partial_{\mu}(g^{\alpha\sigma}\partial_{\beta}{{\b t}^{2}}_{(\sigma\alpha)} 
+ {\tilde t}^{2(\alpha\sigma)}\partial_{\beta}g_{\sigma\alpha} 
- {\tilde t}^{(\alpha\sigma)}\partial_{\beta}{\tilde t}_{(\sigma\alpha)})
-{\tilde t}^{(\alpha\sigma)}(2\partial_{\beta}{\b t}_{(\sigma\mu)} - \partial_{\sigma}{\b t}_{(\mu\beta)})   \nonu
\A \A \hspace{1.5cm} 
+g^{\beta\mu}\partial_{\alpha} \{ g^{\alpha\sigma}(2\partial_{\beta}{{\b t}^{2}}_{(\sigma\mu)} - \partial_{\sigma}
{{\b t}^{2}}_{(\mu\beta)}) 
+ {\tilde t}^{2(\alpha\sigma)}(2\partial_{\beta}g_{\sigma\mu} - \partial_{\sigma}g_{\mu\beta}) \}   \nonu
\A \A \hspace{1.5cm}
+ 2\partial^{\alpha}g_{\lambda\mu}g^{\beta\lambda}(2\partial^{\mu}{{\b t}^{2}}_{(\alpha\beta)}-\partial_{\alpha}
{{\b t}^{2}}_{(\beta\rho)}g^{\mu\rho}) 
+ 2{\tilde t}^{2(\lambda\rho)}\partial_{\lambda}g_{\sigma\mu}g^{\beta\mu}
(2\partial^{\sigma}g_{\beta\rho}-\partial_{\rho}g_{\alpha\beta})                            \nonu
\A \A \hspace{1.5cm}
- 2{\tilde t}^{(\lambda\rho)}\partial_{\lambda}g_{\sigma\mu}g^{\alpha\sigma}
(2\partial^{\mu}{\b t}_{(\rho\alpha)}-\partial_{\rho}{\b t}_{(\alpha\beta)}g^{\beta\mu}) 
+ \partial^{\rho}{\b t}_{(\sigma\mu)}g^{\beta\mu}(2\partial^{\sigma}{{\b t}}_{(\beta\rho)}-\partial_{\rho}
{{\b t}}_{(\alpha\beta)}g^{\alpha\sigma})                                                      \nonu
\A \A \hspace{1.5cm}
+ {\tilde t}^{(\alpha\sigma)}{\tilde t}^{(\lambda\rho)} \{ \partial^{\beta}g_{\lambda\sigma}
\partial_{\beta}g_{\rho\alpha} + 2\partial_{\lambda}g_{\sigma\mu}g^{\mu\beta}
( \partial_{\alpha}g_{\beta\rho}- \partial_{\rho}g_{\alpha\beta}) \}                          \nonu
\A \A \hspace{1.5cm}
- 2{\tilde t}^{(\lambda\rho)}\partial_{\lambda}{\b t}_{(\sigma\mu)}g^{\beta\mu}
(2\partial^{\sigma}g_{\beta\rho}-\partial_{\rho}g_{\alpha\beta}g^{\alpha\sigma})               \nonu
\A \A \hspace{1.5cm}
- \partial^{\rho}g_{\sigma\alpha}g^{\sigma\alpha}(2\partial^{\mu}{{\b t}^{2}}_{(\rho\mu)} - \partial_{\rho}
{{\b t}^{2}}_{(\mu\beta)}g^{\beta\mu}) 
- \partial^{\rho}{{\b t}^{2}}_{(\sigma\alpha)}(2\partial^{\mu}g_{\rho\mu}g^{\sigma\alpha} - \partial^{\lambda}
{g}_{\mu\beta}g^{\mu\beta})                                                           \nonu
\A \A \hspace{1.5cm}
- {\tilde t}^{2(\lambda\rho)}\partial_{\lambda}g_{\sigma\alpha}g^{\sigma\alpha}
(2\partial^{\mu}g_{\rho\mu} - \partial_{\rho}g_{\mu\beta}g^{\mu\beta}) 
- {\tilde t}^{2(\alpha\sigma)}\partial^{\rho}g_{\sigma\alpha}
(2\partial^{\mu}g_{\rho\mu} - \partial^{\rho}g_{\mu\beta}g^{\mu\beta})  \nonu
\A \A \hspace{1.5cm}
- {\tilde t}^{(\lambda\rho)}\partial_{\lambda}g_{\sigma\alpha}g^{\alpha\sigma}
(2\partial^{\mu}{\b t}_{(\rho\mu)}-\partial_{\rho}{\b t}_{(\mu\beta)}g^{\beta\mu}) 
- {\tilde t}^{(\alpha\sigma)}\partial^{\rho}t_{(\sigma\alpha)}
(2\partial^{\mu}{g}_{\rho\mu}-\partial_{\rho}{g}_{\mu\beta}g^{\beta\mu})  \nonu
\A \A \hspace{1.5cm}
+ {g}^{\alpha\sigma}\partial^{\rho}{\tilde t}_{(\sigma\alpha)}
(2\partial^{\mu}{\b t}_{(\rho\mu)} - \partial_{\rho}{\b t}_{(\mu\beta)}g^{\beta\mu}) 
+  {\tilde t}^{(\alpha\sigma)}{\tilde t}^{(\lambda\rho)}\partial_{\lambda}g_{\sigma\alpha}g^{\alpha\sigma}
(2\partial^{\mu}{g}_{\rho\mu} - \partial_{\rho}{g}_{\mu\beta}g^{\beta\mu})      \nonu
\A \A \hspace{1.5cm}
- {\tilde t}^{(\lambda\rho)}\partial_{\lambda}{\b t}_{(\sigma\alpha)}g^{\alpha\sigma}
(2\partial^{\mu}{g}_{\rho\mu} - \partial_{\rho}{g}_{\mu\beta}g^{\beta\mu}) 
- {\tilde t}^{(\alpha\sigma)}\partial^{\rho}{g}_{\sigma\alpha}
(2\partial^{\mu}{\b t}_{(\rho\mu)} - \partial_{\rho}{\b t}_{(\mu\beta)}g^{\beta\mu})     \nonu
\A \A \hspace{1.5cm}
+ {1 \over 2}{\tilde t}^{2(\beta\mu)}  \{ g^{\alpha\sigma}\partial_{\mu}\partial_{\beta}{\b t}_{(\alpha\sigma)}  \nonu
\A \A \hspace{1.5cm} 
- \partial^{\sigma}\partial_{\beta}{\b t}_{(\sigma\mu)} 
+ \partial_{\mu}{\tilde t}^{(\alpha\sigma)} \partial_{\beta}g_{\sigma\alpha}
- \partial_{\mu}g^{\alpha\sigma}\partial_{\beta}{\b t}_{(\sigma\alpha)}            \nonu
\A \A \hspace{1.5cm} 
+ \partial_{\alpha}g^{\alpha\sigma}\partial_{\beta}{\b t}_{(\sigma\mu)} 
- \partial_{\alpha}{\tilde t}^{(\alpha\sigma)} \partial_{\beta}g_{\sigma\mu}     \nonu
\A \A \hspace{1.5cm} 
+ 2\partial^{\rho}{\b t}_{(\sigma\mu)} \partial^{\sigma}g_{\beta\rho}
- 2g^{\alpha\sigma}\partial_{\lambda}{\b t}_{(\sigma\mu)} \partial^{\lambda}g_{\alpha\beta}  \nonu
\A \A \hspace{1.5cm} 
+ g^{\alpha\sigma}g^{\lambda\rho} \partial_{\mu}{\b t}_{(\lambda\sigma)} \partial^{\beta}g_{\rho\alpha} 
- 2g^{\alpha\sigma}\partial^{\rho}{\b t}_{(\sigma\alpha)} \partial_{\beta}g_{\rho\mu}  \nonu
\A \A \hspace{1.5cm} 
+ g^{\alpha\sigma}\partial_{\lambda}{\b t}_{(\sigma\alpha)} \partial^{\lambda}g_{\mu\beta} \}   \nonu
\A \A \hspace{1.5cm} 
- {1 \over 2}{\tilde t}^{(\beta\mu)}  \{  \partial_{\mu}(g^{\alpha\sigma}\partial_{\beta}{{\b t}^{2}}_{(\sigma\alpha)} 
- {\tilde t}^{(\alpha\sigma)}\partial_{\beta}{\tilde t}_{(\sigma\alpha)})
+ \partial_{\mu}{\tilde t}^{2(\alpha\sigma)}\partial_{\beta}{g}_{\sigma\alpha}    \nonu
\A \A \hspace{1.5cm} 
+ \partial_{\alpha}  \{  g^{\alpha\sigma}(2\partial_{\beta}{{\b t}^{2}}_{(\sigma\mu)} - \partial_{\sigma}
{{\b t}^{2}}_{\mu\beta}) 
-{\tilde t}^{(\alpha\sigma)}(2\partial_{\beta}{\b t}_{(\sigma\mu)} - \partial_{\sigma}{\b t}_{(\mu\beta)}  \} 
+ \partial_{\alpha}{\tilde t}^{2(\alpha\sigma)}(2\partial_{\beta}g_{\sigma\mu} - \partial_{\sigma}g_{\mu\beta}) \nonu
\A \A \hspace{1.5cm}
+ 2\partial^{\alpha}g_{\lambda\mu}g^{\beta\lambda}(2\partial^{\mu}{{\b t}^{2}}_{(\alpha\beta)} - \partial_{\alpha}
{{\b t}^{2}}_{(\beta\rho)}g^{\mu\rho}) 
+ 2{\tilde t}^{2(\lambda\rho)}\partial_{\lambda}g_{\sigma\mu}g^{\beta\mu}
(2\partial^{\sigma}g_{\beta\rho}-\partial_{\rho}g_{\alpha\beta})             \nonu
\A \A \hspace{1.5cm}
- 2{\tilde t}^{(\lambda\rho)}\partial_{\lambda}g_{\sigma\mu}g^{\alpha\sigma}
(2\partial^{\mu}{\b t}_{(\rho\alpha)}-\partial_{\rho}{\b t}_{(\alpha\beta)}g^{\beta\mu}) 
+ \partial^{\rho}{\b t}_{(\sigma\mu)}g^{\beta\mu}(2\partial^{\sigma}{{\b t}}_{(\beta\rho)}-\partial_{\rho}
{{\b t}}_{(\alpha\beta)}g^{\alpha\sigma})             \nonu
\A \A \hspace{1.5cm}
+ {\tilde t}^{(\alpha\sigma)}{\tilde t}^{(\lambda\rho)} \{ \partial^{\beta}g_{\lambda\sigma}
\partial_{\beta}g_{\rho\alpha} + 2\partial_{\lambda}g_{\sigma\mu}g^{\mu\beta}
( \partial_{\alpha}g_{\beta\rho}-  \partial_{\rho}g_{\alpha\beta}) \}             \nonu
\A \A \hspace{1.5cm}
- 2{\tilde t}^{(\lambda\rho)}\partial_{\lambda}{\b t}_{(\sigma\mu)}g^{\beta\mu}
(2\partial^{\sigma}g_{\beta\rho}-\partial_{\rho}g_{\alpha\beta}g^{\alpha\sigma})     \nonu
\A \A \hspace{1.5cm}
- \partial^{\rho}g_{\sigma\alpha}g^{\sigma\alpha}(2\partial^{\mu}{{\b t}^{2}}_{(\rho\mu)} - \partial_{\rho}
{{\b t}^{2}}_{(\mu\beta)}g^{\beta\mu}) 
- \partial^{\rho}{{\b t}^{2}}_{(\sigma\alpha)}(2\partial^{\mu}g_{\rho\mu}g^{\sigma\alpha} - \partial^{\lambda}
{g}_{\mu\beta}g^{\mu\beta})  \nonu
\A \A \hspace{1.5cm}
- {\tilde t}^{2(\lambda\rho)}\partial_{\lambda}g_{\sigma\alpha}g^{\sigma\alpha}
(2\partial^{\mu}g_{\rho\mu} - \partial_{\rho}g_{\mu\beta}g^{\mu\beta}) 
- {\tilde t}^{2(\alpha\sigma)}\partial^{\rho}g_{\sigma\alpha}
(2\partial^{\mu}g_{\rho\mu} - \partial^{\rho}g_{\mu\beta}g^{\mu\beta})  \nonu
\A \A \hspace{1.5cm}
- {\tilde t}^{(\lambda\rho)}\partial_{\lambda}g_{\sigma\alpha}g^{\alpha\sigma}
(2\partial^{\mu}{\b t}_{(\rho\mu)}-\partial_{\rho}{\b t}_{(\mu\beta)}g^{\beta\mu}) 
- {\tilde t}^{(\alpha\sigma)}\partial^{\rho}t_{(\sigma\alpha)}
(2\partial^{\mu}{g}_{\rho\mu}-\partial_{\rho}{g}_{\mu\beta}g^{\beta\mu})  \nonu
\A \A \hspace{1.5cm}
+ {g}^{\alpha\sigma}\partial^{\rho}{\tilde t}_{(\sigma\alpha)}
(2\partial^{\mu}{\b t}_{(\rho\mu)} - \partial_{\rho}{\b t}_{(\mu\beta)}g^{\beta\mu}) 
+  {\tilde t}^{(\alpha\sigma)}{\tilde t}^{(\lambda\rho)}\partial_{\lambda}g_{\sigma\alpha}g^{\alpha\sigma}
(2\partial^{\mu}{g}_{\rho\mu} - \partial_{\rho}{g}_{\mu\beta}g^{\beta\mu})      \nonu
\A \A \hspace{1.5cm}
- {\tilde t}^{(\lambda\rho)}\partial_{\lambda}{\b t}_{(\sigma\alpha)}g^{\alpha\sigma}
(2\partial^{\mu}{g}_{\rho\mu} - \partial_{\rho}{g}_{\mu\beta}g^{\beta\mu}) 
- {\tilde t}^{(\alpha\sigma)}\partial^{\rho}{g}_{\sigma\alpha}
(2\partial^{\mu}{\b t}_{(\rho\mu)} - \partial_{\rho}{\b t}_{(\mu\beta)}g^{\beta\mu})    \}    \nonu
\A \A \hspace{1.5cm}
+ \partial^{\beta} \{  ({\tilde t}^{2(\alpha\sigma)}\partial_{\beta}{\b t}_{(\sigma\alpha)} 
- {\tilde t}^{(\alpha\sigma)}\partial_{\beta}{\b t}{^2}_{(\sigma\alpha)})  \}  \nonu
\A \A \hspace{1.5cm}
- g^{\beta\mu}\partial_{\alpha} \{ {\tilde t}^{2(\alpha\sigma)}(2\partial_{\beta}{\b t}_{(\sigma\mu)} 
-\partial_{\sigma}{\b t}_{(\mu\beta)}) 
- {\tilde t}^{(\alpha\sigma)}(2\partial_{\beta}{{\b t}^2}_{(\sigma\mu)}
-\partial_{\beta}{{\b t}^{2}}_{(\mu\beta)})  \}           \nonu
\A \A \hspace{1.5cm}
+ {1 \over 2} \{ g^{\alpha\sigma}(\partial_{\lambda}g_{\sigma\mu}+\partial_{\mu}g_{\lambda\sigma}
-\partial_{\sigma}g_{\mu\lambda}) {\tilde t}^{2(\lambda\rho)}g^{\beta\mu}(\partial_{\beta}{\b t}_{(\rho\alpha)}+
\partial_{\alpha}{\b t}_{(\beta\rho)}-\partial_{\rho}{\b t}_{(\alpha\beta)})    \nonu
\A \A \hspace{1.5cm}
+ g^{\alpha\sigma}(\partial_{\lambda}{\b t}_{(\sigma\mu)}+\partial_{\mu}g_{\lambda\sigma}
-\partial_{\sigma}{\b t}_{(\mu\lambda)}) {\tilde t}^{2(\lambda\rho)}{\b t}^{(\lambda\rho)}g^{\beta\mu}
(\partial_{\beta}g_{\rho\alpha}+\partial_{\alpha}g_{\beta\rho}-\partial_{\rho}g_{\alpha\beta})       \nonu
\A \A \hspace{1.5cm}
- g^{\alpha\sigma}(\partial_{\lambda}g_{\sigma\mu}+\partial_{\mu}g_{\lambda\sigma}
-\partial_{\sigma}g_{\mu\lambda}){\tilde t}^{(\lambda\rho)}g^{\beta\mu}
(\partial_{\beta}{{\b t}^{2}}_{(\rho\alpha)}+\partial_{\alpha}{{\b t}^{2}}_{(\beta\rho)}-
\partial_{\rho}{{\b t}^{2}}_{(\alpha\beta)})       \nonu
\A \A \hspace{1.5cm}
- g^{\alpha\sigma}(\partial_{\lambda}{{\b t}^{2}}_{(\sigma\mu)}+\partial_{\mu}{{\b t}^{2}}_{(\lambda\sigma)}
-\partial_{\sigma}{{\b t}^{2}}_{(\mu\lambda)}) {\tilde t}^{(\lambda\rho)}g^{\beta\mu}(\partial_{\beta}g_{\rho\alpha}+
\partial_{\alpha}g_{\beta\rho}-\partial_{\rho}g_{\alpha\beta})       \nonu
\A \A \hspace{1.5cm}
+ {\tilde t}^{(\alpha\sigma)}(\partial_{\lambda}g_{\sigma\mu}+\partial_{\mu}g_{\lambda\sigma}
-\partial_{\sigma}g_{\mu\lambda}){\tilde t}^{(\lambda\rho)}g^{\beta\mu}(\partial_{\beta}{\b t}_{(\rho\alpha)}+
\partial_{\alpha}{\b t}_{(\beta\rho)}-\partial_{\rho}{\b t}_{(\alpha\beta)})     \nonu
\A \A \hspace{1.5cm}
- g^{\alpha\sigma}(\partial_{\lambda}{\b t}_{(\sigma\mu)}+\partial_{\mu}g_{\lambda\sigma}
-\partial_{\sigma}{\b t}_{(\mu\lambda)}){\tilde t}^{(\lambda\rho)}g^{\beta\mu}(\partial_{\beta}{\b t}_{(\rho\alpha)}+
\partial_{\alpha}{\b t}_{(\beta\rho)}-\partial_{\rho}{\b t}_{(\alpha\beta)})     \nonu
\A \A \hspace{1.5cm}
- {\tilde t}^{(\alpha\sigma)}(\partial_{\lambda}{\b t}_{(\sigma\mu)}+\partial_{\mu}g_{\lambda\sigma}
-\partial_{\sigma}{\b t}_{(\mu\lambda)})g^{\lambda\rho}g^{\beta\mu}(\partial_{\beta}{\b t}_{(\rho\alpha)}+
\partial_{\alpha}{\b t}_{(\beta\rho)}-\partial_{\rho}{\b t}_{(\alpha\beta)})          \nonu
\A \A \hspace{1.5cm}  
- g^{\alpha\sigma}\partial_{\lambda}g_{\sigma\alpha}{\tilde t}^{2\lambda\rho}g^{\beta\mu}
(2\partial_{\beta}{\b t}_{(\rho\mu)}-\partial_{\rho}{\b t}_{(\mu\beta)})    \nonu
\A \A \hspace{1.5cm}
+ g^{\alpha\sigma}\partial_{\lambda}{\b t}_{(\sigma\alpha)} {\tilde t}^{2(\lambda\rho)}
(2\partial_{\beta}g_{\rho\mu}-\partial_{\rho}g_{\mu\beta})       \nonu
\A \A \hspace{1.5cm}
+ g^{\alpha\sigma}\partial_{\lambda}g_{\sigma\alpha}{\tilde t}^{(\lambda\rho)}g^{\beta\mu}
(2\partial_{\beta}{{\b t}^{2}}_{(\rho\mu)}-\partial_{\rho}{{\b t}^{2}}_{(\mu\beta)})       \nonu
\A \A \hspace{1.5cm}
+ g^{\alpha\sigma}\partial_{\lambda}{{\b t}^{2}}_{(\sigma\mu)}{\tilde t}^{(\lambda\rho)}
g^{\beta\mu}(2\partial_{\beta}g_{\rho\mu}-\partial_{\rho}g_{\mu\beta})       \nonu
\A \A \hspace{1.5cm}
- {\tilde t}^{(\alpha\sigma)}\partial_{\lambda}g_{\sigma\alpha}{\tilde t}^{(\lambda\rho)}
g^{\beta\mu}(2\partial_{\beta}{\b t}_{(\rho\mu)}-\partial_{\rho}{\b t}_{(\mu\beta)})     \nonu
\A \A \hspace{1.5cm}
+ g^{\alpha\sigma}\partial_{\lambda}{\b t}_{(\sigma\alpha)}{\tilde t}^{(\lambda\rho)}
g^{\beta\mu}(2\partial_{\beta}{\b t}_{(\rho\mu)}-\partial_{\rho}{\b t}_{(\mu\beta)})     \nonu
\A \A \hspace{1.5cm}
+ {\tilde t}^{(\alpha\sigma)}(\partial_{\lambda}{\b t}_{(\sigma\alpha)}+\partial_{\alpha}g_{\lambda\sigma}
-\partial_{\sigma}{\b t}_{(\alpha\lambda)})g^{\lambda\rho}g^{\beta\mu}(\partial_{\beta}{\b t}_{(\rho\mu)}+
\partial_{\mu}{\b t}_{(\beta\rho)}-\partial_{\rho}{\b t}_{(\mu\beta)})   \}  \nonu
\A \A \hspace{1.5cm}
- {\tilde t}^{(\beta\mu)} \{  {1 \over 2} \{  \partial_{\mu}{\tilde t}^{(\alpha\sigma)}
\partial_{\beta}{{\b t}^{2}}_{(\sigma\alpha)} - \partial_{\mu}{{\tilde t}^{2}}_{(\alpha\sigma)}
\partial_{\beta}{\b t}_{(\sigma\alpha)} +
{\tilde t}^{(\alpha\sigma)}\partial_{\mu}\partial_{\beta}{{\b t}^{2}}_{(\sigma\alpha)}            \nonu
\A \A \hspace{1.5cm}
- \partial_{\alpha}{\tilde t}^{(\alpha\sigma)}(2\partial_{\beta}{{\b t}^{2}}_{(\sigma\mu)}
-\partial_{\sigma}{{\b t}^{2}}_{(\mu\beta)}) + \partial_{\alpha}{{\tilde t}^{2}}_{(\alpha\sigma)}
(2\partial_{\beta}{\b t}_{(\sigma\mu)}- \partial_{\sigma}{\b t}_{(\mu\beta)})                      \nonu
\A \A \hspace{1.5cm}
- {\tilde t}^{(\alpha\sigma)}\partial_{\alpha}(2\partial_{\beta}{{\b t}^{2}}_{(\sigma\mu)}
- \partial_{\sigma}{{\b t}^{2}}_{(\mu\beta)})  \} \nonu
\A \A \hspace{1.5cm}
+{1 \over 4} \{  g^{\alpha\sigma}(\partial_{\lambda}{\b t}_{(\sigma\mu)}+\partial_{\mu}{\b t}_{(\lambda\sigma)}
-\partial_{\sigma}{\b t}_{(\mu\lambda)})g^{\lambda\rho} (\partial_{\beta}{{\b t}^{2}}_{(\rho\alpha)}
+\partial_{\alpha}{{\b t}^{2}}_{(\beta\rho)}-\partial_{\rho}{{\b t}^{2}}_{(\alpha\beta)})       \nonu
\A \A \hspace{1.5cm}
- g^{\alpha\sigma}(\partial_{\lambda}{\b t}_{(\sigma\mu)}+\partial_{\mu}g_{\lambda\sigma}
-\partial_{\sigma}{\b t}_{(\mu\lambda)}){\tilde t}^{(\lambda\rho)}(\partial_{\beta}{{\b t}}_{(\rho\alpha)}
+\partial_{\alpha}{{\b t}}_{(\beta\rho)}-\partial_{\rho}{{\b t}}_{(\alpha\beta)})               \nonu
\A \A \hspace{1.5cm}
+ g^{\alpha\sigma}(\partial_{\lambda}{{\b t}^{2}}_{(\sigma\mu)}+\partial_{\mu}{{\b t}^{2}}_{(\lambda\sigma)}
-\partial_{\sigma}{{\b t}^{2}}_{(\mu\lambda)})g^{\lambda\rho} (\partial_{\beta}{{\b t}}_{(\rho\alpha)}
+\partial_{\alpha}{{\b t}}_{(\beta\rho)}-\partial_{\rho}{{\b t}}_{(\alpha\beta)})                 \nonu
\A \A \hspace{1.5cm}
- {\tilde t}^{(\alpha\sigma)}(\partial_{\lambda}{{\b t}}_{(\sigma\mu)}+\partial_{\mu}{{\b t}}_{(\lambda\sigma)}
-\partial_{\sigma}{{\b t}}_{(\mu\lambda)})g^{\lambda\rho} (\partial_{\beta}{{\b t}}_{(\rho\alpha)}
+\partial_{\alpha}{{\b t}}_{(\beta\rho)}-\partial_{\rho}{{\b t}}_{(\alpha\beta)})        \nonu
\A \A \hspace{1.5cm}
- g^{\alpha\sigma}\partial_{\lambda}{\b t}_{(\sigma\alpha)}g^{\lambda\rho}
(2\partial_{\beta}{{\b t}^{2}}_{(\rho\mu)}-\partial_{\rho}{{\b t}^{2}}_{(\mu\beta)})         \nonu
\A \A \hspace{1.5cm}
+g^{\alpha\sigma}\partial_{\lambda}{\b t}_{(\sigma\alpha)}{\tilde t}^{(\lambda\rho)}
(2\partial_{\beta}{{\b t}}_{(\rho\mu)}-\partial_{\rho}{{\b t}}_{(\mu\beta)})                 \nonu
\A \A \hspace{1.5cm}
- g^{\alpha\sigma}\partial_{\lambda}{{\b t}^{2}}_{(\sigma\alpha)}g^{\lambda\rho}
(2\partial_{\beta}{{\b t}}_{(\rho\mu)}-\partial_{\rho}{{\b t}}_{(\mu\beta)})                \nonu
\A \A \hspace{1.5cm}
+ {\tilde t}^{(\alpha\sigma)}\partial_{\lambda}{\b t}_{(\sigma\alpha)}
g^{\lambda\rho}(2\partial_{\beta}{{\b t}}_{(\rho\mu)}-\partial_{\rho}{{\b t}}_{(\mu\beta)}) \}   \}  \nonu
\A \A \hspace{1.5cm}
+ g^{\beta\mu} \{ {1 \over 2} \{ \partial_{\mu}{\tilde t}^{2(\alpha\sigma)}\partial_{\beta}{{\b t}^{2}}_{(\sigma\alpha)}
+ {\tilde t}^{2(\alpha\sigma)} \partial_{\mu}\partial_{\beta}{{\b t}^{2}}_{(\sigma\alpha)}              \nonu
\A \A \hspace{1.5cm}
- \partial_{\alpha}{\tilde t}^{2(\alpha\sigma)}(2\partial_{\beta}{{\b t}^{2}}_{(\sigma\mu)}
-\partial_{\sigma}{{\b t}^{2}}_{(\mu\beta)})                                              \nonu
\A \A \hspace{1.5cm}
- {\tilde t}^{2(\alpha\sigma)} \partial_{\alpha}(2\partial_{\beta}{{\b t}^{2}}_{(\sigma\mu)}
-\partial_{\sigma}{{\b t}^{2}}_{(\mu\beta)})   \}                                           \nonu
\A \A \hspace{1.5cm}
+ {1 \over 4} \{  g^{\alpha\sigma}(\partial_{\lambda}{{\b t}^{2}}_{(\sigma\mu)}+\partial_{\mu}{{\b t}^{2}}_{(\lambda\sigma)}
-\partial_{\sigma}{{\b t}^{2}}_{(\mu\lambda)})g^{\lambda\rho}(\partial_{\beta}{{\b t}^{2}}_{(\rho\alpha)}+
\partial_{\alpha}{{\b t}^{2}}_{(\beta\rho)}-\partial_{\rho}{{\b t}^{2}}_{(\alpha\beta)})        \nonu
\A \A \hspace{1.5cm}
- g^{\alpha\sigma}(\partial_{\lambda}{\b t}_{(\sigma\mu)}+\partial_{\mu}{\b t}_{(\lambda\sigma)}
-\partial_{\sigma}{\b t}_{(\mu\lambda)}){\tilde t}^{(\lambda\rho)}(\partial_{\beta}{{\b t}^{2}}_{(\rho\alpha)}+
\partial_{\alpha}{{\b t}^{2}}_{(\beta\rho)}-\partial_{\rho}{{\b t}^{2}}_{(\alpha\beta)})          \nonu
\A \A \hspace{1.5cm}
+ g^{\alpha\sigma}(\partial_{\lambda}{\b t}_{(\sigma\mu)}+\partial_{\mu}{\b t}_{(\lambda\sigma)}
-\partial_{\sigma}{\b t}_{(\mu\lambda)}){\tilde t}^{2(\lambda\rho)}(\partial_{\beta}{{\b t}}_{(\rho\alpha)}+
\partial_{\alpha}{{\b t}}_{(\beta\rho)}-\partial_{\rho}{{\b t}}_{(\alpha\beta)})                  \nonu
\A \A \hspace{1.5cm}
- g^{\alpha\sigma}(\partial_{\lambda}{{\b t}^{2}}_{(\sigma\mu)}+\partial_{\mu}{{\b t}^{2}}_{(\lambda\sigma)}
-\partial_{\sigma}{{\b t}^{2}}_{(\mu\lambda)}){\tilde t}^{(\lambda\rho)}(\partial_{\beta}{{\b t}}_{(\rho\alpha)}+
\partial_{\alpha}{{\b t}}_{(\beta\rho)}-\partial_{\rho}{{\b t}}_{(\alpha\beta)})                  \nonu
\A \A \hspace{1.5cm}
- {\tilde t}^{(\alpha\sigma)}(\partial_{\lambda}{\b t}_{(\sigma\mu)}+\partial_{\mu}{\b t}_{(\lambda\sigma)}
-\partial_{\sigma}{\b t}_{(\mu\lambda)})g^{\lambda\rho}(\partial_{\beta}{{\b t}^{2}}_{(\rho\alpha)}+
\partial_{\alpha}{{\b t}^{2}}_{(\beta\rho)}-\partial_{\rho}{{\b t}^{2}}_{(\alpha\beta)})             \nonu
\A \A \hspace{1.5cm}
+ {\tilde t}^{(\alpha\sigma)}(\partial_{\lambda}{\b t}_{(\sigma\mu)}+\partial_{\mu}{\b t}_{(\lambda\sigma)}
-\partial_{\sigma}{\b t}_{(\mu\lambda)}){\tilde t}^{(\lambda\rho)}(\partial_{\beta}{{\b t}}_{(\rho\alpha)}+
\partial_{\alpha}{{\b t}}_{(\beta\rho)}-\partial_{\rho}{{\b t}}_{(\alpha\beta)})                 \nonu
\A \A \hspace{1.5cm}
+  {\tilde t}^{2(\alpha\sigma)}(\partial_{\lambda}{\b t}_{(\sigma\mu)}+\partial_{\mu}{\b t}_{(\lambda\sigma)}
-\partial_{\sigma}{\b t}_{(\mu\lambda)})g^{\lambda\rho}(\partial_{\beta}{{\b t}}_{(\rho\alpha)}+
\partial_{\alpha}{{\b t}}_{(\beta\rho)}-\partial_{\rho}{{\b t}}_{(\alpha\beta)})                        \nonu
\A \A \hspace{1.5cm}
- g^{\alpha\sigma}\partial_{\lambda}{{\b t}^{2}}_{(\sigma\mu)}g^{\lambda\rho}
(2\partial_{\beta}{{\b t}^{2}}_{(\rho\mu)}-\partial_{\rho}{{\b t}^{2}}_{(\mu\beta)})                  \nonu
\A \A \hspace{1.5cm}
+ g^{\alpha\sigma}\partial_{\lambda}{\b t}_{(\sigma\alpha)}{\tilde t}^{(\lambda\rho)}
(2\partial_{\beta}{{\b t}^{2}}_{(\rho\mu)}-\partial_{\rho}{{\b t}^{2}}_{(\mu\beta)})                    \nonu
\A \A \hspace{1.5cm}
- g^{\alpha\sigma}\partial_{\lambda}{\b t}_{(\sigma\alpha)}{\tilde t}^{2(\lambda\rho)}
(2\partial_{\beta}{{\b t}}_{(\rho\mu)}-\partial_{\rho}{{\b t}}_{(\mu\beta)})                            \nonu
\A \A \hspace{1.5cm}
+ g^{\alpha\sigma}\partial_{\lambda}{{\b t}^{2}}_{(\sigma\mu)}{\tilde t}^{(\lambda\rho)}
(2\partial_{\beta}{{\b t}}_{(\rho\mu)}-\partial_{\rho}{{\b t}}_{(\mu\beta)})                            \nonu
\A \A \hspace{1.5cm}
+  {\tilde t}^{(\alpha\sigma)}\partial_{\lambda}{\b t}_{(\sigma\alpha)}g^{\lambda\rho}
(2\partial_{\beta}{{\b t}^{2}}_{(\rho\mu)}-\partial_{\rho}{{\b t}^{2}}_{(\mu\beta)})               \nonu
\A \A \hspace{1.5cm}
- {\tilde t}^{(\alpha\sigma)}\partial_{\lambda}{\b t}_{(\sigma\alpha)}{\tilde t}^{(\lambda\rho)}
(2\partial_{\beta}{{\b t}}_{(\rho\mu)}-\partial_{\rho}{{\b t}}_{(\mu\beta)})                         \nonu
\A \A \hspace{1.5cm}
- {\tilde t}^{2(\alpha\sigma)}\partial_{\lambda}{\b t}_{(\sigma\alpha)}g^{\lambda\rho}
(2\partial_{\beta}{{\b t}}_{(\rho\mu)}-\partial_{\rho}{{\b t}}_{(\mu\beta)})     \}  \} \big],           \nonu
\A \A \hspace{1.5cm}
\label{2dSGM-2}
\ea
where $R_{\mu\nu\rho\sigma}$, $R_{\mu\nu}$ and $R$ are the curvature tensors of Riemann space and
$\vert w_{V-A} \vert = \{ 1+{t^{a}}_{a}+{1 \over 2}({t^{a}}_{a}{t^{b}}_{b}-{t^{a}}_{b}{t^{b}}_{a}) \}$ 
is V-A model in  two dimensional flat space.  Note that the result is still preliminary, for the 
multiplication by $\vert w_{V-A} \vert$ factorized in (\ref{2dSGM-2}) should be expanded 
in the power series in $t$.      \\

{\bf 3.2 SGM in the spin connection formalism}       \\
Next we perform the similar arguments in the spin connection formalism 
for the sake of the comparison. 
The spin connection $Q{^{ab}}_{\mu}$ 
and the curvature tensor $\Omega{^{ab}}_{\mu \nu}$ 
in SGM space-time are as follows; 
\ba
\A \A Q_{ab \mu} 
= {1 \over 2} (w{_{[a}}^{\rho} \partial_{\mid \mu \mid} w_{b] \rho} 
- w{_{[a}}^{\rho} \partial_{\mid \rho \mid} w_{b] \mu} 
- w{_{[a}}^{\rho} w{_{b]}}^{\sigma} 
w_{c \mu} \partial_{\rho} w{^c}_{\sigma}), \\
\A \A \Omega{^{ab}}_{\mu \nu} 
= \partial_{[\mu} Q{^{ab}}_{\nu]} + Q{^a}_{c[\mu} Q{^{cb}}_{\nu]}. 
\ea
The scalar curvature $\Omega$ of SGM space-time is defined by 
$\Omega = w{^a}_{\mu} w{^b}_{\nu} \Omega{^{ab}}_{\mu \nu}$. 
Let us express the spin connection $Q{^{ab}}_{\mu}$ 
in two dimensional space-time in terms of 
$e{^a}_{\mu}$ and $t{^a}_{\mu}$ as 
\begin{equation}
Q_{ab \mu} = \omega_{ab \mu}[e] 
+ T^{(1)}_{ab \mu} + T^{(2)}_{ab \mu} 
+ T^{(3)}_{ab \mu}, 
\end{equation}
where $\omega_{ab \mu}[e]$ 
is the Ricci rotation coefficients of GR, 
and $T^{(1)}_{ab \mu}$, $T^{(2)}_{ab \mu}$ 
and $T^{(3)}_{ab \mu}$ are defined as 
\ba
T^{(1)}_{ab \mu} 
= \A \A {1 \over 2} (e{_{[a}}^{\rho} \partial_{\mid \mu \mid} t_{b] \rho} 
- t{^{\rho}}_{[a} \partial_{\mid \mu \mid} e_{b] \rho} 
- e{_{[a}}^{\rho} \partial_{\mid \rho \mid} t_{b] \mu} 
+ t{^{\rho}}_{[a} \partial_{\mid \rho \mid} e_{b] \mu} 
\nonu
\A \A - e{_{[a}}^{\rho} e{_{b]}}^{\sigma} 
e_{c \mu} \partial_{\rho} t{^c}_{\sigma} 
+ e{_{[a}}^{[\rho} t{^{\sigma]}}_{b]} 
e_{c \mu} \partial_{\rho} e{^c}_{\sigma} 
- e{_{[a}}^{\rho} e{_{b]}}^{\sigma} 
t_{c \mu} \partial_{\rho} e{^c}_{\sigma}), 
\\
T^{(2)}_{ab \mu} 
= \A \A {1 \over 2} (- t{^{\rho}}_{[a} \partial_{\mid \mu \mid} t_{b] \rho} 
+ t{^{\rho}}_{[a} t{^{\sigma}}_{\mid \rho} 
\partial_{\mu \mid} e_{b] \sigma} 
+ t{^{\rho}}_{[a} \partial_{\mid \rho \mid} t_{b] \mu} 
- t{^{\rho}}_{[a} t{^{\sigma}}_{\mid \rho} 
\partial_{\sigma \mid} e_{b] \mu} 
\nonu
\A \A + e{_{[a}}^{[\rho} t{^{\sigma]}}_{b]} 
e_{c \mu} \partial_{\rho} t{^c}_{\sigma} 
- e{_{[a}}^{\rho} e{_{b]}}^{\sigma} 
t_{c \mu} \partial_{\rho} t{^c}_{\sigma} 
- e{_{[a}}^{[\rho} t{^{\mid \sigma \mid}}_{b]} 
t{^{\lambda]}}_{\sigma} e_{c \mu} 
\partial_{\rho} e{^c}_{\lambda} 
\nonu
\A \A - t{^{\rho}}_{[a} t{^{\sigma}}_{b]} 
e_{c \mu} \partial_{\rho} e{^c}_{\sigma} 
+ e{_{[a}}^{[\rho} t{^{\sigma]}}_{b]} t_{c \mu} 
\partial_{\rho} e{^c}_{\sigma}), 
\\
T^{(3)}_{ab \mu} 
= \A \A {1 \over 2} (t{^{\rho}}_{[a} t{^{\sigma}}_{\mid \rho} 
\partial_{\mu \mid} t_{b] \sigma} 
- t{^{\rho}}_{[a} t{^{\sigma}}_{\mid \rho} 
\partial_{\sigma \mid} t_{b] \mu} 
\nonu
\A \A 
- e{_{[a}}^{[\rho} t{^{\mid \sigma \mid}}_{b]} 
t{^{\lambda]}}_{\sigma} e_{c \mu} 
\partial_{\rho} t{^c}_{\lambda} 
- t{^{\rho}}_{[a} t{^{\sigma}}_{b]} 
e_{c \mu} \partial_{\rho} t{^c}_{\sigma} 
+ e{_{[a}}^{[\rho} t{^{\sigma]}}_{b]} t_{c \mu} 
\partial_{\rho} t{^c}_{\sigma}), 
\ea
where $t{^{\mu}}_a = e{_b}^{\mu} e{_a}^{\nu} t{^b}_{\nu}$. 
Note that $T^{(1)}_{ab \mu}$ and $T^{(2)}_{ab \mu}$ 
can be written by using the spin connection 
$\omega{^{ab}}_{\mu}[e]$ of GR as 
\ba
T^{(1)}_{ab \mu} 
= \A \A e{_{[a}}^{\rho} \hat D_{\mid \mu \mid} t_{b] \rho} 
+ {1 \over 4} e{_{[a}}^{\rho} e{_{b]}}^{\sigma} 
\partial_{\dot\mu} t_{[\dot\rho \dot\sigma]}, 
\\
T^{(2)}_{ab \mu} 
= \A \A - t{^{\rho}}_{\sigma} 
e{_{[a}}^{\sigma} \hat D_{\mid \mu \mid} t_{b] \rho} 
+ {1 \over 2} e{_{[a}}^{\rho} e{_{b]}}^{\sigma} 
t_{c \rho} \hat D_{\mu} t{^c}_{\sigma} 
\nonu
\A \A 
- {1 \over 2} e{_{[a}}^{\rho} e{_{b]}}^{\sigma} 
\partial_{\rho} (t_{c \mu} t{^c}_{\sigma}) 
- {1 \over 2} t{^{\rho}}_{\lambda} 
e{_{[a}}^{\lambda} e{_{b]}}^{\sigma} 
\partial_{\dot\mu} t_{[\dot\rho \dot\sigma]}, 
\ea
where $\hat D_{\mu} t_{a \nu} := 
\partial_{\mu} t_{a \nu} 
+ \omega_{ab \mu} t{^{b}}_{\nu}$ 
and $\partial_{\dot\mu} t_{[\dot\rho \dot\sigma]} 
:= \partial_{\mu} t_{[\rho \sigma]} 
+ \partial_{\sigma} t_{(\mu \rho)} 
- \partial_{\rho} t_{(\sigma \mu)}$. 
Then we obtain straightforwardly 
the complete expression of 2 dimensional SGM action(N=1) 
in the spin connection formalism as follows; namely,   \\
\ba
L_{2dSGM} = \A \A 
- {{c^3 \Lambda} \over 16{\pi}G} e \vert w_{V-A} \vert 
\nonu
\A \A
- {c^3 \over 16{\pi}G} e \vert w_{V-A} \vert 
[R - 4 t^{\mu \nu} R_{\mu \nu} 
+ 2 e^{a[\mu} e^{\mid b \mid \nu]} 
(\hat D_{\mu} e{_a}^{\rho}) \hat D_{\nu} t_{b \rho} 
+ D_{\mu} (g^{\mu \rho} g^{\nu \sigma} 
\partial_{\dot\nu} t_{[\dot\rho \dot\sigma]}) 
\nonu
\A \A
+ 2 (t^{\rho \mu} t{^{\nu}}_{\rho} + t^{\rho \nu} t{^{\mu}}_{\rho} 
+ t^{\mu \rho} t{^{\nu}}_{\rho}) R_{\mu \nu} 
+ t^{(\mu \rho)} t^{(\nu \sigma)} R_{\mu \nu \rho \sigma} 
\nonu
\A \A
- (g^{\rho [\mu} g^{\mid \kappa \mid \nu]} g^{\sigma \lambda} 
+ g^{\sigma [\mu} g^{\mid \kappa \mid \nu]} g^{\rho \lambda} 
- g^{\sigma [\mu} g^{\mid \lambda \mid \nu]} g^{\kappa \rho}) 
e{^a}_{\sigma} e{^b}_{\kappa} (\hat D_{\mu} t_{a \rho}) 
\hat D_{\nu} t_{b \lambda} 
\nonu
\A \A
+ g^{\rho [\mu} g^{\mid \kappa \mid \nu]} g^{\sigma \lambda} 
e{^a}_{\sigma} (\hat D_{\mu} t_{a \rho}) 
\partial_{\dot\nu} t_{[\dot\lambda \dot\kappa]} 
+ {1 \over 4} g^{\rho [\mu} g^{\mid \kappa \mid \nu]} g^{\sigma \lambda} 
(\partial_{\dot\mu} t_{[\dot\rho \dot\sigma]}) 
\partial_{\dot\nu} t_{[\dot\lambda \dot\kappa]} 
\nonu
\A \A
- 2 (g^{\rho [\mu} e^{\mid b \mid \nu]} \hat D_{\mu} e^{a \sigma} 
+ e^{c [\mu} e^{\mid b \mid \nu]} e^{a \sigma} \hat D_{\mu} e{_c}^{\rho} 
\nonu
\A \A
+ e^{c [\mu} e^{\mid a \mid \nu]} e^{b \rho} \hat D_{\mu} e{_c}^{\sigma} 
- e^{b [\mu} e^{\mid a \mid \nu]} e^{c \rho} \hat D_{\mu} e{_c}^{\sigma}) 
t_{a \rho} \hat D_{\nu} t_{b \sigma} 
\nonu
\A \A
- e^{a [\mu} g^{\mid \rho \mid \nu]} (\hat D_{\mu} e{_a}^{\sigma}) 
t_{b [\rho} \hat D_{\mid \nu \mid} t{^b}_{\sigma]} 
- g^{\rho \nu} e^{a \mu} e^{b \sigma} (\hat D_{\mu} e{_b}^{\lambda}) 
t_{a \lambda} \partial_{\dot\nu} t_{[\dot\rho \dot\sigma]} 
\nonu
\A \A
- D_{\mu} \{ g^{\rho [\mu} g^{\mid \sigma \mid \nu]} 
\partial_{\rho}(t_{a \nu} t{^a}_{\sigma}) 
+ 2 g^{\rho [\mu} t^{\nu] \sigma} 
\partial_{\dot\nu} t_{[\dot\rho \dot\sigma]} \} 
\nonu
\A \A
- e^{[a \mid \mu \mid} (\hat D_{\mu} e^{b] \nu}) 
\{ t{^{\rho}}_a t{^{\sigma}}_{\rho} \partial_{[\nu} t_{\mid b \mid \sigma]} 
- (e{_a}^{\rho} t{^{\lambda}}_b t{^{\sigma}}_{\lambda} e_{c \nu} 
+ {1 \over 2} t{^{\rho}}_a t{^{\sigma}}_b e_{c \nu} 
- e{_a}^{\rho} t{^{\sigma}}_b t_{c \nu}) 
\partial_{[\rho} t{^c}_{\sigma]} \} 
\nonu
\A \A
+ e^{a [\mu} t^{\nu] b} 
\{ 2(\hat D_{\mu} t{^{\rho}}_{[a}) \hat D_{\mid \nu \mid} t_{b] \rho} 
- e{_{[a}}^{\rho} (\hat D_{\mid \mu \mid} e{_{b]}}^{\sigma}) 
t_{c [\rho} \partial_{\mid \nu \mid} t{^c}_{\sigma]} \} 
\nonu
\A \A
- g^{\rho [\mu} e{_a}^{\nu]} e{_b}^{\sigma} 
(\hat D_{\mu} t^{ab}) \partial_{[\rho} 
(t_{\mid c \nu \mid} t{^c}_{\sigma]}) 
- e^{c [\mu} e{_a}^{\nu]} e{_{[c}}^{\lambda} e{_{b]}}^{\sigma} 
t{^{\rho}}_{\lambda} (\hat D_{\mu} t^{ab}) 
\partial_{\dot\nu} t_{[\dot\rho \dot\sigma]} 
\nonu
\A \A
+ (2 e^{a [\mu} t^{\mid \lambda b \mid} t{^{\nu]}}_{\lambda} 
+ t^{[\mu \mid a \mid} t^{\nu] b}) 
\{ (\hat D_{\mu} e{_{[a}}^{\rho}) \partial_{\mid \nu \mid} t_{b] \rho} 
+ {1 \over 2} e{_{[a}}^{\rho} (\hat D_{\mid \mu \mid} e{_{b]}}^{\sigma}) 
\partial_{\dot\nu} t_{[\dot\rho \dot\sigma]} 
\nonu
\A \A
+ {1 \over 2} e{_a}^{\rho} e{_b}^{\sigma} \partial_{\mu} 
\partial_{\dot\nu} t_{[\dot\rho \dot\sigma]} \} 
\nonu
\A \A
+ 2 g^{\rho [\mu} g^{\mid \sigma \mid \nu]} t{^{\lambda}}_{\sigma} 
(\hat D_{\mu} t_{a \rho}) \hat D_{\nu} t{^a}_{\lambda} 
\nonu
\A \A
- 2 (g^{\rho [\mu} e^{\mid b \mid \nu]} t^{\lambda a} 
- e^{a [\mu} e^{\mid b \mid \nu]} t^{\lambda \rho} 
+ e^{a [\mu} t^{\mid \lambda \mid \nu]} e^{b \rho}) 
(\hat D_{\mu} t_{a \rho}) \hat D_{\nu} t_{b \lambda} 
\nonu
\A \A
- \{ (g^{\rho [\mu} t^{\nu] b} - e^{b [\mu} t^{\nu] \rho}) e^{a \lambda} 
- (e^{a [\mu} t^{\nu] b} - e^{b [\mu} t^{\nu] a}) g^{\rho \lambda} 
\nonu
\A \A
+ (e^{a [\mu} t^{\nu] \lambda} - g^{\lambda [\mu} t^{\nu] a}) e^{b \rho} \} 
(\hat D_{\mu} t_{a \rho}) \hat D_{\nu} t_{b \lambda} 
\nonu
\A \A
+ 2 (g^{\rho [\mu} g^{\mid \sigma \mid \nu]} e^{a \lambda} 
- e^{a [\mu} g^{\mid \sigma \mid \nu]} g^{\rho \lambda}) 
\{ (\hat D_{\mu} t_{a \rho}) 
t_{c [\lambda} \hat D_{\mid \nu \mid} t{^c}_{\sigma]} 
- (\hat D_{\mu} t_{a \rho}) \partial_{[\lambda} 
(t_{\mid b \nu \mid} t{^b}_{\sigma]}) \} 
\nonu
\A \A
- \{ (g^{\rho [\mu} t^{\nu] \kappa} - g^{\kappa [\mu} t^{\nu] \rho}) 
e^{a \lambda} 
- (e^{a [\mu} t^{\nu] \kappa} - g^{\kappa [\mu} t^{\nu] a}) 
g^{\rho \lambda} \} (\hat D_{\mu} t_{a \rho}) 
\partial_{\dot\nu} t_{[\dot\lambda \dot\kappa]} 
\nonu
\A \A
- (g^{\rho [\mu} g^{\mid \kappa \mid \nu]} t^{\lambda a} 
- g^{\rho [\mu} t^{\mid \lambda \mid \nu]} e^{a \kappa} 
- e^{a [\mu} g^{\mid \kappa \mid \nu]} t^{\lambda \rho} 
+ e^{a [\mu} t^{\mid \lambda \mid \nu]} g^{\rho \kappa} 
\nonu
\A \A
+ g^{\lambda [\mu} e^{\mid a \mid \nu]} t^{\rho \kappa} 
- g^{\lambda [\mu} t^{\mid \rho \mid \nu]} e^{a \kappa}) 
(\hat D_{\mu} t_{a \rho}) \partial_{\dot\nu} t_{[\dot\lambda \dot\kappa]} 
\nonu
\A \A
- {1 \over 2} g^{\lambda [\mu} g^{\mid \sigma \mid \nu]} g^{\kappa \rho} 
\{ t_{a [\rho} (\hat D_{\mid \mu \mid} t{^a}_{\sigma]}) 
- \partial_{[\rho} (t_{\mid a \mu \mid} t{^a}_{\sigma]}) \} 
\partial_{\dot\nu} t_{[\dot\lambda \dot\kappa]} 
\nonu
\A \A
- {1 \over 2} (g^{\lambda [\mu} g^{\mid \sigma \mid \nu]} t^{\rho \kappa} 
- g^{\lambda [\mu} g^{\mid \alpha \mid \nu]} g^{\kappa \sigma} 
t{^{\rho}}_{\alpha}) (\partial_{\dot\mu} t_{[\dot\rho \dot\sigma]}) 
\partial_{\dot\nu} t_{[\dot\lambda \dot\kappa]} 
\nonu
\A \A
- {1 \over 4} (g^{\rho [\mu} t^{\nu] \kappa} - g^{\kappa [\mu} t^{\nu] \rho}) 
g^{\sigma \lambda} 
(\partial_{\dot\mu} t_{[\dot\rho \dot\sigma]}) 
\partial_{\dot\nu} t_{[\dot\lambda \dot\kappa]} 
\nonu
\A \A
+ {1 \over 2} D_{\mu} \{ e^{a [\mu} e^{\mid b \mid \nu]} 
\{ 2 t{^{\rho}}_a t{^{\sigma}}_{\rho} \partial_{[\nu} t_{\mid b \mid \sigma]} 
- (2 e{_a}^{\rho} t{^{\lambda}}_b t{^{\sigma}}_{\lambda} e_{c \nu} 
+ t{^{\rho}}_a t{^{\sigma}}_b e_{c \nu} 
- 2 e{_a}^{\rho} t{^{\sigma}}_b t_{c \nu}) 
\partial_{[\rho} t{^c}_{\sigma]} \} \}
\nonu
\A \A
+ D_{\mu} \{ g^{\rho [\mu} t^{\nu] \sigma} 
\partial_{[\rho} (t_{\mid a \nu \mid} t{^a}_{\sigma]}) 
+ (g^{\lambda [\mu} t^{\nu] \sigma} - g^{\sigma [\mu} t^{\nu] \lambda}) 
t{^{\rho}}_{\lambda} \partial_{\dot\nu} t_{[\dot\rho \dot\sigma]} \} 
\nonu
\A \A
+ {1 \over 4} (e^{a [\mu} e{_d}^{\nu]} e^{b \kappa} 
- e^{b [\mu} e{_d}^{\nu]} e^{a \kappa}) 
\partial_{\mu} t{^d}_{\kappa} 
\{ 2 t{^{\rho}}_a t{^{\sigma}}_{\rho} \partial_{[\nu} t_{\mid b \mid \sigma]} 
\nonu
\A \A
- (2 e{_a}^{\rho} t{^{\lambda}}_b t{^{\sigma}}_{\lambda} e_{c \nu} 
+ t{^{\rho}}_a t{^{\sigma}}_b e_{c \nu} 
- 2 e{_a}^{\rho} t{^{\sigma}}_b t_{c \nu}) 
\partial_{[\rho} t{^c}_{\sigma]} \} 
\nonu
\A \A
- (e^{a [\mu} t{^{\nu]}}_{\rho} t^{\rho b} 
- e^{b [\mu} t{^{\nu]}}_{\rho} t^{\rho a} + t^{[\mu \mid a \mid} t^{\nu] b}) 
\nonu
\A \A
\times \{ e{_c}^{\rho} e{_a}^{\sigma} (\partial_{\mu} t{^c}_{\sigma}) 
\partial_{[\nu} t_{\mid b \mid \rho]} 
- {1 \over 2} e{_a}^{\rho} e{_b}^{\sigma} (\partial_{\mu} t_{c \nu}) 
\partial_{\rho} t{^c}_{\sigma} 
- e{_c}^{\rho} e{_a}^{\lambda} e{_b}^{\sigma} (\partial_{\mu} t{^c}_{\lambda})\partial_{[\sigma} 
t_{\mid \nu \mid \rho]} \} 
\nonu
\A \A
- g^{\sigma [\mu} g^{\mid \kappa \mid \nu]} 
t{^{\rho}}_{\sigma} t{^{\lambda}}_{\kappa} 
(\partial_{\mu} t_{a \rho}) \partial_{\nu} t{^a}_{\lambda} 
\nonu
\A \A
+ (g^{\sigma [\mu} e^{\mid b \mid \nu]} 
t{^{\rho}}_{\sigma} t^{\lambda a} 
- e^{a [\mu} e^{\mid b \mid \nu]} 
t{^{\rho}}_{\sigma} t^{\lambda \sigma} 
+ e^{a [\mu} g^{\mid \kappa \mid \nu]} 
t^{\rho b} t{^{\lambda}}_{\kappa}) 
(\partial_{\mu} t_{a \rho}) \partial_{\nu} t_{b \lambda} 
\nonu
\A \A
- {1 \over 2} (g^{\sigma [\mu} g^{\mid \kappa \mid \nu]} e^{a \lambda} 
- e^{a [\mu} g^{\mid \kappa \mid \nu]} g^{\sigma \lambda}) 
t{^{\rho}}_{\sigma} (\partial_{\mu} t_{a \rho}) 
\{ t_{d [\lambda} \partial_{\mid \nu \mid} t{^d}_{\kappa]} 
- \partial_{[\lambda} (t_{\mid d \nu \mid} t{^d}_{\kappa]}) \} 
\nonu
\A \A
+ {1 \over 2} (g^{\sigma [\mu} g^{\mid \kappa \mid \nu]} e^{a \alpha} 
- g^{\sigma [\mu} g^{\mid \alpha \mid \nu]} e^{a \kappa} 
- e^{a [\mu} g^{\mid \kappa \mid \nu]} g^{\sigma \alpha} 
+ e^{a [\mu} g^{\mid \alpha \mid \nu]} g^{\sigma \kappa}) 
t{^{\rho}}_{\sigma} t{^{\lambda}}_{\alpha} 
(\partial_{\mu} t_{a \rho}) \partial_{\dot\nu} t_{[\dot\lambda \dot\kappa]} 
\nonu
\A \A
- {1 \over 2} (g^{\rho [\mu} e^{\mid a \mid \nu]} g^{\sigma \kappa} 
- g^{\rho [\mu} g^{\mid \kappa \mid \nu]} e^{a \sigma}) 
\{ t_{b [\rho} \partial_{\mid \mu \mid} t{^b}_{\sigma]} 
- \partial_{[\rho} (t_{\mid b \mu \mid} t{^b}_{\sigma]}) \} 
t^{\lambda}_{\kappa} \partial_{\nu} t_{a \lambda} 
\nonu
\A \A
+ {1 \over 4} g^{\rho [\mu} g^{\mid \kappa \mid \nu]} g^{\sigma \lambda} 
\{ t_{a [\rho} \partial_{\mid \mu \mid} t{^a}_{\sigma]} 
- \partial_{[\rho} (t_{\mid a \mu \mid} t{^a}_{\sigma]}) \} 
\{ t_{b [\lambda} \partial_{\mid \nu \mid} t{^b}_{\kappa]} 
- \partial_{[\lambda} (t_{\mid b \nu \mid} t{^b}_{\kappa]}) \} 
\nonu
\A \A
- {1 \over 4} (g^{\rho [\mu} g^{\mid \kappa \mid \nu]} g^{\sigma \alpha} 
- g^{\rho [\mu} g^{\mid \alpha \mid \nu]} g^{\sigma \kappa}) 
\{ t_{a [\rho} \partial_{\mid \mu \mid} t{^a}_{\sigma]} 
- \partial_{[\rho} (t_{\mid a \mu \mid} t{^a}_{\sigma]}) \} 
t{^{\lambda}}_{\alpha} \partial_{\dot\nu} t_{[\dot\lambda \dot\kappa]} 
\nonu
\A \A
+ {1 \over 2} (t^{\rho [\mu} e^{\mid a \mid \nu]} t^{\lambda \sigma} 
- t^{\rho [\mu} t^{\lambda \mid \nu]} e^{a \sigma} 
- g^{\sigma [\mu} e^{\mid a \mid \nu]} t^{\rho \kappa} t{^{\lambda}}_{\kappa} 
+ g^{\sigma [\mu} t^{\lambda \mid \nu]} t^{\rho a}) 
(\partial_{\dot\mu} t_{[\dot\rho \dot\sigma]}) 
\partial_{\nu} t_{a \lambda} 
\nonu
\A \A
- {1 \over 4} (g^{\alpha [\mu} g^{\mid \kappa \mid \nu]} g^{\sigma \lambda} 
- g^{\sigma [\mu} g^{\mid \kappa \mid \nu]} g^{\alpha \lambda}) 
t{^{\rho}}_{\alpha} (\partial_{\dot\mu} t_{[\dot\rho \dot\sigma]}) 
\{ t_{a [\lambda} \partial_{\mid \nu \mid} t{^a}_{\kappa]} 
- \partial_{[\lambda} (t_{\mid a \nu \mid} t{^a}_{\kappa]}) \} 
\nonu
\A \A
+ {1 \over 4} (g^{\alpha [\mu} g^{\mid \kappa \mid \nu]} g^{\sigma \beta} 
- g^{\alpha [\mu} g^{\mid \beta \mid \nu]} g^{\sigma \kappa} 
- g^{\sigma [\mu} g^{\mid \kappa \mid \nu]} g^{\alpha \beta} 
+ g^{\sigma [\mu} g^{\mid \beta \mid \nu]} g^{\alpha \kappa}) 
t{^{\rho}}_{\alpha} t{^{\lambda}}_{\beta} 
(\partial_{\dot\mu} t_{[\dot\rho \dot\sigma]}) 
\partial_{\dot\nu} t_{[\dot\lambda \dot\kappa]} 
\nonu
\A \A
+ \{ g^{\rho [\mu} e^{\mid b \mid \nu]} t^{\lambda a} t{^{\kappa}}_{\lambda} 
(\partial_{\mu} t_{a \rho}) \partial_{[\nu} t_{\mid b \mid \kappa]} 
- g^{\rho [\mu} t^{\mid \lambda \mid \nu]} t{^{\kappa}}_{\lambda} 
(\partial_{\mu} t_{a \rho}) \partial_{[\nu} t{^a}_{\kappa]} \} 
\nonu
\A \A
- (g^{\rho [\mu} t^{\mid \sigma \mid \nu]} e^{a \lambda} 
- g^{\rho [\mu} g^{\mid \lambda \mid \nu]} t^{\sigma a} 
- e^{a [\mu} t^{\mid \sigma \mid \nu]} g^{\rho \lambda} 
+ e^{a [\mu} g^{\mid \lambda \mid \nu]} t^{\sigma \rho}) 
\nonu
\A \A
\times \{ e_{d \nu} t{^{\kappa}}_{\sigma} (\partial_{\mu} t_{a \rho}) 
\partial_{[\lambda} t{^d}_{\kappa]} 
- t_{d \nu} (\partial_{\mu} t_{a \rho}) \partial_{[\lambda} t{^d}_{\sigma]} \} 
\nonu
\A \A
- (g^{\rho [\mu} e^{\mid b \mid \nu]} t^{\lambda a} 
- e^{a [\mu} e^{\mid b \mid \nu]} t^{\lambda \rho}) 
e_{d \nu} t{^{\kappa}}_b (\partial_{\mu} t_{a \rho}) 
\partial_{[\lambda} t{^d}_{\kappa]} 
\nonu
\A \A
+ {1 \over 2} (g^{\rho [\mu} e^{\mid a \mid \nu]} t^{\lambda \sigma} 
- g^{\rho [\mu} t^{\mid \lambda \mid \nu]} e^{a \sigma}) 
t{^{\kappa}}_{\lambda} (\partial_{\dot\mu} t_{[\dot\rho \dot\sigma]}) 
\partial_{[\nu} t_{\mid a \mid \kappa]} 
\nonu
\A \A
- {1 \over 2} \{ (g^{\rho [\mu} t^{\mid \alpha \mid \nu]} g^{\lambda \sigma} 
- g^{\rho [\mu} g^{\mid \lambda \mid \nu]} t^{\alpha \sigma}) 
t{^{\kappa}}_{\alpha} e_{a \nu} 
+ g^{\rho [\mu} t^{\mid \kappa \mid \nu]} t^{\lambda \sigma} e_{a \nu} 
\nonu
\A \A
- (g^{\rho [\mu} t^{\mid \kappa \mid \nu]} g^{\lambda \sigma} 
- g^{\rho [\mu} g^{\mid \lambda \mid \nu]} t^{\kappa \sigma}) t_{a \nu} \} 
(\partial_{\dot\mu} t_{[\dot\rho \dot\sigma]}) 
\partial_{[\lambda} t{^a}_{\kappa]} 
\nonu
\A \A
+ \{ (g^{\rho [\mu} t{^{\nu]}}_{\sigma} t^{\sigma b} 
- e^{b [\mu} t{^{\nu]}}_{\sigma} t^{\sigma \rho} 
+ t^{[\mu \mid \rho \mid} t^{\nu] b}) e^{a \lambda} 
\nonu
\A \A
- (e^{a [\mu} t{^{\nu]}}_{\sigma} t^{\sigma b} 
- e^{b [\mu} t{^{\nu]}}_{\sigma} t^{\sigma a} 
+ t^{[\mu \mid a \mid} t^{\nu] b}) g^{\rho \lambda} 
\nonu
\A \A
+ (e^{a [\mu} t{^{\nu]}}_{\sigma} t^{\sigma \lambda} 
- g^{\lambda [\mu} t{^{\nu]}}_{\sigma} t^{\sigma a} 
+ t^{[\mu \mid a \mid} t^{\nu] \lambda}) e{_b}^{\rho} \} 
(\partial_{\mu} t_{a \rho}) \partial_{\nu} t_{b \lambda} 
\nonu
\A \A
- (g^{\rho [\mu} t{^{\nu]}}_{\sigma} t^{\sigma \lambda} 
- g^{\lambda [\mu} t{^{\nu]}}_{\sigma} t^{\sigma \rho} 
+ t^{[\mu \mid \rho \mid} t^{\nu] \lambda}) 
(\partial_{\mu} t_{a \rho}) \partial_{\nu} t{^a}_{\lambda} 
\nonu
\A \A
+ \{ (g^{\rho [\mu} t{^{\nu]}}_{\sigma} t^{\sigma \kappa} 
- g^{\kappa [\mu} t{^{\nu]}}_{\sigma} t^{\sigma \rho} 
+ t^{[\mu \mid \rho \mid} t^{\nu] \kappa}) e^{a \lambda} 
\nonu
\A \A
- (e^{a [\mu} t{^{\nu]}}_{\sigma} t^{\sigma \kappa} 
- g^{\kappa [\mu} t{^{\nu]}}_{\sigma} t^{\sigma a} 
+ t^{[\mu \mid a \mid} t^{\nu] \kappa}) g^{\rho \lambda} \} 
(\partial_{\mu} t_{a \rho}) 
\partial_{\dot\nu} t_{[\dot\lambda \dot\kappa]} 
\nonu
\A \A
+ {1 \over 4} (g^{\rho [\mu} t{^{\nu]}}_{\alpha} t^{\alpha \kappa} 
- g^{\kappa [\mu} t{^{\nu]}}_{\alpha} t^{\alpha \rho} 
+ t^{[\mu \mid \rho \mid} t^{\nu] \kappa}) g^{\sigma \lambda} 
(\partial_{\dot\mu} t_{[\dot\rho \dot\sigma]}) 
\partial_{\dot\nu} t_{[\dot\lambda \dot\kappa]} ], 
\label{2dSGM-spin}
\ea
where $\hat D_{\mu} T_{ab \nu} := 
\partial_{\mu} T_{ab \nu} 
+ \omega_{ac \mu} T{^{c}}_{b \nu} 
+ \omega_{bc \mu} T{_a}{^{c}}_{\nu}$ 
and $D_{\mu} e{_a}^{\nu} := \partial_{\mu} e{_a}^{\nu} 
+ \Gamma^{\nu}_{\lambda \mu} e{_a}^{\lambda}$.    \\
As in the affine connection case  the result is still preliminary, 
for the multiplication by $\vert w_{V-A} \vert$ factorized in (\ref{2dSGM-spin}) should be expanded 
in the power series in $t$.     \\

{\bf 4. Discussions}             \\
We can summerize the fundmental idea as fiollows.
In SGM the tangent spacetime is specified by $(x^{a}, \psi(x))$
and  $w{^a}_{\mu}(x)$ is the unified vierbein(, i.e. only one dynamical field) of SGM spacetime 
which has  a global superGL(4,R) and the ordinary local GL(4,R) transformations.
In Einstein gravity the tangent spacetime is specified by $x^{a}$
and  $e{^a}_{\mu}(x)$  is the vierbein of  Riemann spacetime which has the ordinary GL4,R) transformation.
By the geometrical arguments in each space there appear fundamental invariant actions  
which have the same features in appearance. 
However SGM action in SGM spacetime written in terms of $w{^a}_{\mu}(x)$ is 
unstable(degenerated) against the global new NL SUSY  and (SGM spacetime) 
breaks down spontaneously
to $e{^a}_{\mu}(x)$(Riemann metric)+ $t{^a}_{\mu}(x)$(a specific composite of N-Gfermion matter)
by reducing the initial gravitational (vacuum) energy through  the conversion
to  matter energy. Note that SGM action posesses two inequivalent flat spaces. 
The dynamical degrees of freedom  are  now  $e{^a}_{\mu}(x)$ and  $\psi(x)$
(not the composite $t{^a}_{\mu}(x)$) in the ordinary Riemann(Einstein gravity) spacetime  and
SGM action describes  the ordinary GL(4,R) invariant dynamics of $e{^a}_{\mu}(x)$ and $\psi(x)$.
And (in general as well) it is difficult to eliminate $\psi(x)$ without
introducing a symmetry with a local fermionic arbitrary gauge parameter,
e.g. a local fermionic gauge invariant coupling to SUGRA.      \par
We have shown that contrary to its simple expression (\ref{SGMac}) in unified SGM space-time
the  expansion of SGM action posesses very complicated and rich structures 
describing as a whole(not order by order) gauge invariant  graviton-superon interactions.  \\
The final results after carrying out the multiplication of 
$\vert w_{V-A} \vert$ may be rewritten in a  simpler form up to total derivative terms. 
As mentioned above SGM action is remarkably a free action of  
E-H type action  in unified SGM space-time, the various familiar classical exact solutions of GR may be 
reinterpreted as those of SGM( ${w_{a}}^{\mu}(x)$  and metric  ${s}^{\mu\nu}(x)$) and may have new 
physical meanings for EGRT.           \par
Concerning the abovementioned two inequivalent flat-spaces (i.e. the vacuum of the gravitational energy) 
of SGM action we can interpret them as follows.    \par
SGM action (\ref{SGMac}) written by the vierbein ${w_{a}}^{\mu}(x)$  and metric  ${s}^{\mu\nu}(x)$ 
of SGM space-time is invariant under  (besides the ordinary local GL(4,R)) the general coordinate 
transformation\cite{st2}  with a generalized parameter ${i \kappa \bar{\zeta}{\gamma}^{\mu}\psi(x)}$ 
(originating from the global supertranslation \cite{va} 
$\psi(x)  \rightarrow  \psi(x) + \zeta$ in SGM space-time).  
As proved for E-H action of  GR\cite{wttn}, the energy of SGM action of E-H type is expected to be  positive 
(for positive $\Lambda$).
Regarding the scalar curvature tensor $\Omega$ for the unified metric tensor $s^{\mu\nu}(x)$ as an analogue 
of  the Higgs potential for the Higgs scalar, 
we can observe  that (at least the vacuum of)  SGM action, 
(i.e. SGM-flat $w{^a}_{\mu}(x) \rightarrow {\delta}{^a}_{\nu}$ space-time,)  
which allows Riemann space-time and   has a positive  energy density 
with the positive cosmological constant 
${c^3\Lambda \over 16{\pi}G}$ indicating the spontaneous SUSY bgeaking, 
is unstable (,i.e. degenerates) 
against the  supertransformation  (\ref{newSUSY1/2}) and (\ref{newSUSY2}) with the global spinor parameter $\zeta$ 
in SGM space-time    
and breaks down  spontaneously to Riemann space-time  
\ba
\A \A 
w{^a}_{\mu}(x) = e{^a}_{\mu}(x) + t{^a}_{\mu}(x),
\label{S-Gmetric}
\ea
with N-G fermions $superons$ corresponding to 
\ba
\A \A        { superGL(4,R)  \over  GL(4,R) }.  
\label{e-tEX}
\ea
(Note that SGM-flat space-time allows Riemann space-time.)
Remarkably  the observed Riemann space-time of EGRT and matter(superons) appear  simultaneously from 
(the vacuum of) SGM action by the spontaneous SUSY breaking.                   \par
The analysis of the structures of the vacuum of Riemann-flat space-time (described by N=10 V-A action with 
derivative terms similar to (\ref{b-va}) ) plays an important role 
to linearlize SGM and to derive SM as the low energy effective theory, which remain to be challenged.   
The derivative terms can be rewritten in the tractable form (\ref{b-va})  up to the total derivative terms.  
The linearlization of the flat-space N=1 V-A model  was already carried out\cite{rikuzw}. 
The linearlization of N=2 V-A mdel is extremely important from the physical point of view, 
for it gives a new mechanism  generating a (U(1)) gauge field of the linearlized (effective) theory \cite{ks4}. 
In our case of SGM the algebra( gauge symmetry ) should be changed to 
broken SO(10) SP(broken SUGRA \cite{dzfnf})symmetry by the linearlization  
which is isomorphic to the initial one (\ref{SGMgr}). The systematic and generic arguments on the relation of 
linerar and nonlinear SUSY are already investigated\cite{wb}. 
Recently we have shown that N=1 gauge vector multiplet action with  SUSY breaking Feyet-Iliopolos  term is 
equivalent to N=1 flat space V-A action of NL SUSY\cite{stt1}.  
A U(1) gauge field, though an axial vector field for N=1 case, is expressed by N-G field 
(and its highly  nonlinear self interactions).                                        \par
Finally we just mention the hidden symmetries  characteristic to SGM.  
It is natural to expect that SGM action may be invariant under a certain exchange 
between ${e^{a}}_{\mu}$ and ${t^{a}}_{\mu}$, 
for they contribute equally to the unified SGM vierbein  ${w^{a}}_\mu$ as seen in (\ref{new-w}).
In fact we find, as a simple example, that SGM action is invariant under the following exchange  
of ${e^{a}}_{\mu}$ and ${t^{a}}_{\mu}$\cite{st3} (in 4 dimensional space-time).  

\ba
\A \A {e^{a}}_{\mu}  \longrightarrow  2{t^{a}}_{\mu}, 
{t^{a}}_{\mu} \longrightarrow  {e^{a}}_{\mu} -  {t^{a}}_{\mu},  \nonu
\A \A \hspace{1.5cm} 
{e_{a}}^{\mu} \longrightarrow {e_{a}}^{\mu}, 
\label{e-tEX}
\ea
or in terms of the  metric it can be written as  
\ba
\A \A g_{\mu\nu} \longrightarrow 4{t^{\rho}}_{\mu}t_{\rho\nu}, 
t_{\mu\nu} \longrightarrow 2(t_{\nu\mu}-t_{\rho\mu}{t^{\rho}}_{\nu}),  \nonu
\A \A \hspace{1.5cm} 
g^{\mu\nu} \longrightarrow g^{\mu\nu}, t^{\mu\nu} \longrightarrow  g^{\mu\nu} - t^{\mu\nu}. 
\label{g-tEX}
\ea
This can be generalized to the following form with two real(one complex) global prameters\cite{st3},         \\
\begin{equation}
\pmatrix{e{^a}_{\mu} \cr
         t{^a}_{\mu} \cr
         t{^b}_{\mu} e{_b}^{\nu} t{^a}_{\nu} \cr} 
\rightarrow \pmatrix{
            0 & 2(\alpha + 1) & -2(\alpha^2 - \beta) \cr
            1 & -(2 \alpha + 1) & 2(\alpha^2 - \beta) \cr
            1 & -(3 \alpha + 2) & 2 \alpha(2 \alpha + 1) - 3 \beta + 1 \cr} 
\ \pmatrix{e{^a}_{\mu} \cr
         t{^a}_{\mu} \cr
         t{^b}_{\mu} e{_b}^{\nu} t{^a}_{\nu} \cr}. 
\end{equation}
The physical meaning  of such   symmetries  remains to be studied.         \\   
Also SGM action has $Z_{2}$ symmetry $\psi^{j} \rightarrow -\psi^{j}$.              \\

Besides the composite picture of SGM it is interesting to consider (elementary field) SGM with the extra dimensions 
and their  compactifications. The compactification of ${w^{A}}_M={e^{A}}_M + {t^{A}}_M, (A,M=0,1,..D-1)$  
produces rich spectrum of particles and (hidden) internal symmetries and may give a new framework for 
the unification of space-time and matter.          \\

The work of M. Tsuda is supported in part by  High-Tech research program of
Saitama Institute of Technology.
\newpage

%
\newcommand{\NP}[1]{{\it Nucl.\ Phys.\ }{\bf #1}}
\newcommand{\PL}[1]{{\it Phys.\ Lett.\ }{\bf #1}}
\newcommand{\CMP}[1]{{\it Commun.\ Math.\ Phys.\ }{\bf #1}}
\newcommand{\MPL}[1]{{\it Mod.\ Phys.\ Lett.\ }{\bf #1}}
\newcommand{\IJMP}[1]{{\it Int.\ J. Mod.\ Phys.\ }{\bf #1}}
\newcommand{\PR}[1]{{\it Phys.\ Rev.\ }{\bf #1}}
\newcommand{\PRL}[1]{{\it Phys.\ Rev.\ Lett.\ }{\bf #1}}
\newcommand{\PTP}[1]{{\it Prog.\ Theor.\ Phys.\ }{\bf #1}}
\newcommand{\PTPS}[1]{{\it Prog.\ Theor.\ Phys.\ Suppl.\ }{\bf #1}}
\newcommand{\AP}[1]{{\it Ann.\ Phys.\ }{\bf #1}}

\end{document}